\documentclass[iop,apj,twocolumn,numberedappendix]{emulateapj}
\usepackage{amsmath,bm}
\usepackage{epstopdf}

\begin{document}

\title{The Type Ia Supernova Color-Magnitude Relation and Host Galaxy Dust:  \\ A Simple Hierarchical Bayesian Model}
\shorttitle{Hierarchical Bayesian SN Ia}
\author{{Kaisey~S.~Mandel}\altaffilmark{1,6}, {Daniel~M.~Scolnic}\altaffilmark{2}, {Hikmatali~Shariff}\altaffilmark{3}, {Ryan~J.~Foley}\altaffilmark{4} and {Robert~P.~Kirshner}\altaffilmark{1,5} }
\altaffiltext{1}{Harvard-Smithsonian Center for Astrophysics, 60 Garden St., Cambridge, MA 02138, USA}
\altaffiltext{2}{Hubble Fellow, KICP Fellow, Department of Physics, The University of Chicago, Chicago, IL 60627, USA}
\altaffiltext{3}{Astrophysics Group, Physics Department \& Imperial Centre for Inference and Cosmology, Imperial College London, Prince Consort Rd, London SW7 2AZ, UK}
\altaffiltext{4}{Department of Astronomy and Astrophysics, University of California, Santa Cruz, CA 95064, USA}
\altaffiltext{5}{Gordon and Betty Moore Foundation, 1661 Page Mill Road, Palo Alto, CA 94028, USA}
\altaffiltext{6}{kmandel@cfa.harvard.edu}

\slugcomment{Accepted for publication in ApJ}

\begin{abstract}
Conventional Type Ia supernova (SN Ia) cosmology analyses currently use a simplistic linear regression of magnitude versus color and light curve shape, which does not model  intrinsic SN Ia variations and host galaxy dust as physically distinct effects, resulting in low color-magnitude slopes.  We construct a probabilistic generative model for the dusty distribution of extinguished absolute magnitudes and apparent colors as the convolution of a intrinsic SN Ia color-magnitude distribution and a host galaxy dust reddening-extinction distribution.  If the intrinsic color-magnitude ($M_B$ vs. $B-V$) slope $\beta_\text{int}$ differs from the host galaxy dust law $R_B$, this convolution results in a specific curve of mean extinguished absolute magnitude vs. apparent color.   The derivative of this curve smoothly transitions from $\beta_\text{int}$ in the blue tail to $R_B$ in the red tail of the apparent color distribution.  The conventional linear fit approximates this effective curve near the average apparent color, resulting in an apparent slope $\beta_\text{app}$ between $\beta_\text{int}$ and $R_B$.   We incorporate these effects into a hierarchical Bayesian statistical model for SN Ia light curve measurements, and analyze a dataset of SALT2 optical light curve fits of 248 nearby SN Ia at $z < 0.10$.  The conventional linear fit obtains $\beta_\text{app} \approx 3$.   Our model finds a $\beta_\text{int} = 2.3 \pm 0.3$ and a distinct dust law of $R_B = 3.8 \pm 0.3$, consistent with the average for Milky Way dust, while correcting a systematic distance bias of $\sim 0.10$ mag in the tails of the apparent color distribution.  Finally, we extend our model to examine the SN Ia luminosity-host mass dependence in terms of intrinsic and dust components.
\end{abstract}

\section{Introduction}

Type Ia supernova (SN Ia) rest-frame optical light curves have been used as cosmological distance indicators to trace the history of cosmic expansion, detect cosmic acceleration \citep{riess98,perlmutter99}, and to constrain the equation-of-state parameter $w$ of dark energy  \citep{garnavich98b,wood-vasey07, astier06, kowalski08, hicken09b, kessler09, freedman09, amanullah10,conley11,sullivan11,rest14,scolnic14b, betoule14}.   Determining supernova distances with high precision and small systematic error is essential to accurate  constraints on the cosmic expansion history and the properties of dark energy.   

Inferring peak optical absolute magnitudes of SN Ia from distance-independent measures such as their light curve shapes and colors underpins the evidence for cosmic acceleration.  Empirical studies show that SN Ia with broader, slower declining optical light curves are more luminous (``broader-brighter'') and that SN Ia with redder colors are dimmer.  But the ``redder-dimmer'' color-luminosity relation widely used in cosmological SN Ia analyses masks the fact that it has two separate physical origins.  An intrinsic correlation arises from the physics of exploding white dwarfs while interstellar dust in the host galaxy also makes supernovae appear dimmer and redder.  The conventional approach of fitting a single (usually linear) function for luminosity vs. color, for a given light curve shape, is too simple.  This leads to considerable uncertainty regarding the physical interpretation of the color-luminosity distribution of SN Ia, the confounding of extrinsic host galaxy dust reddening with the intrinsic color variations of SN Ia, and the proper way to use SN Ia color measurements to estimate accurate photometric distances.  In this paper, we present a new probabilistic model describing the apparent SN Ia color-magnitude distribution as arising from the combination of intrinsic color-luminosity variations and host galaxy dust reddening and extinction, and apply this to SN Ia data to determine the characteristics of these physical components.

Cosmological analyses of high-$z$ SN Ia data depend on empirical correlations originally observed in samples of nearby low-$z$ SN Ia \citep{hamuy96_29sne, riess99, jha06, hicken09a, contreras10, stritzinger11, hicken12}.  Light curve fitting methods, including MLCS \citep{riess96, riess98, jha07}, SALT2 \citep{guy07, guy10}, SNooPy \citep{burns11}, and \textsc{BayeSN} \citep{mandel11},  all make use of the optical luminosity-light curve width correlation  \citep{phillips93,hamuy96, phillips99}.   However, current approaches conceptually differ on how measured apparent colors are used to infer the SN Ia luminosities and thus estimate the photometric distance.  Methods such as MLCS, SNooPy, and \textsc{BayeSN}  explicitly model the intrinsic SN Ia light curves and the effects of host galaxy dust extinction as separate components.  However, the most popular tool for fitting cosmological SN Ia light curves is currently SALT2, a spectral template model that does not attempt to separate intrinsic SN Ia variations from host galaxy dust effects.

A longstanding puzzle in the analysis of SN Ia light curves is the nature of their apparent color and brightness variations.  In principle, they comprise color and luminosity variations intrinsic to the supernovae, as well as reddening and extinction by interstellar dust along the line of sight in their host galaxies.   However, the fact that astronomers only observe the combination of these effects poses a challenging inference problem.  The function of dust absorption over wavelength \citep[e.g. CCM,][]{ccm89} is typically parameterized by the ratio of total to selective extinction, $R_V = A_V/(A_B - A_V)$.  This ratio normally has an average value of 3.1 for interstellar dust in the Milky Way (MW) Galaxy, although it can vary between 2.1 and 5.8 \citep{draine03}.   \citet{schlafly16} find that the MW extinction curve is fairly uniform: with a narrow spread $\sigma(R_V) \approx 0.18$.  Similar extinction curves have been found in external galaxies; for example, \citet{finkelman08, finkelman10} found average values of $R_V \approx 2.8$.

Early analyses of SN Ia data estimated unphysically low values of $R_V \lesssim 1$ \citep[c.f.][for a review]{branch92}, although these analyses did not take into account empirical correlations between the luminosity, color and light curve shape of the events.  \citet*{riess96b} used the first MLCS \citep*{riess96} method to fit SN Ia optical $BVRI$ light curves and, by minimizing the Hubble diagram scatter, they found a dust $R_V = 2.6 \pm 0.3$, consistent with the Milky Way average.  They noted that the failure to properly account for intrinsic color-luminosity correlations would cause estimates of the dust extinction-reddening ratio $R_V$ to be biased low.    

Extending the wavelength range of the observations from the optical to the rest-frame NIR improves the constraints on the dust law and thus helps disentangle intrinsic color variation from dust reddening.  For several nearby, highly reddened (peak apparent $B-V \gg 0.6$) SN Ia with optical and NIR light curves, $R_V$ can be fit precisely and unusually low values of 1.5-1.8 have been reported \citep{krisciunas07, elias-rosa06, elias-rosa07, wangx08}.  While the origin of these apparently low $R_V$ values is poorly understood \citep{lwang05, goobar08, phillips13,johansson14, amanullah15}, these extremely red and dim objects are not present in the cosmological sample due to selection effects and cuts ($B-V < 0.3$).  

\citet{freedman09} constructed a SN Ia Hubble diagram using rest-frame $i$-band magnitudes and $B-V$ colors.  By minimizing the Hubble residuals $\chi^2$, they estimated $R_V \approx 1.74 \pm 0.27$, and suggested that either dust in SN Ia host galaxies has a substantially different extinction law than Milky Way dust, or that there is significant SN Ia intrinsic color-luminosity dispersion, independent of the light curve shape.  The latter hypothesis was supported by \citet{folatelli10} to explain a discrepancy found when analyzing of the nearby Carnegie Supernova Project \citep[CSP,][]{contreras10} SN Ia sample.  When examining the optical-NIR colors they inferred $R_V =  3.2 \pm 0.4$ when the extremely red objects ($E(B-V) \gtrsim 1$) were excluded.  However, when minimizing the Hubble diagram dispersion, they find low values of $R_V \approx 1-2$.  Recent analyses of larger optical-NIR nearby samples found that the majority of SN Ia Ia with low reddening appear extinguished by dust with $R_V$ closer to $3$ \citep{mandel11, phillips12, burns14}. 

Another strategy for separating intrinsic color variation from dust effects is to measure spectral features that correlate with the intrinsic photometric properties of SN Ia.  For example, \citet{foleykasen11} have correlated the velocity of the Si II $\lambda$6355 absorption feature with the peak intrinsic $B-V$ color, and found $R_V \approx 2.5$.  As they controlled for optical decline rate, this is evidence for independent intrinsic color variations.  \citet{chotard11} modeled the components of the apparent SN Ia spectroscopic and photometric variations depending upon Si II and Ca II H\&K equivalent widths, finding that the remainder is well described  by a CCM dust reddening law with $R_V = 2.8 \pm 0.3$, consistent with the MW value.  However, the vast majority of the high-$z$ SN Ia observations currently used for cosmological analysis consist of rest-frame optical photometry and lack the high-quality spectra or rest-frame NIR data used by the above analyses.

In the following subsections, we describe the conventional method, the Tripp formula, for modeling correlations between SN Ia magnitude, color and light curve shapes from optical light curve data, currently used to estimate cosmological distances.  \textsc{SALT2mu}  is a generalization of this method to account for scatter around the Tripp formula \citep{marriner11}.  We comment on the drawbacks of these approaches, primarily because they are inadequate for properly accounting for the physically distinct factors of intrinsic SN Ia variation and host galaxy dust underlying the data.  We introduce a new statistical model, \textsc{Simple-BayeSN}, that we have developed to address these shortcomings.  \textsc{Simple-BayeSN} analyzes the peak apparent magnitude, apparent color, and light curve shape obtained from light curve fits to the SN Ia photometric time series.  It models the SN Ia data as arising from a probabilistic generative process combining intrinsic SN Ia variations, host galaxy dust effects, and measurement error.   \textsc{Simple-BayeSN} uses a hierarchical Bayesian framework to fit the SN Ia data on the Hubble Diagram, while coherently estimating the parameters driving the underlying effects.

\subsection{The Tripp Formula}

Conventional cosmological SN Ia analysis \citep[e.g.][]{betoule14, rest14} currently proceeds by fitting primarily rest-frame optical light curve data to obtain estimates of peak apparent magnitude $m_s$, apparent color $c_s$ and light curve shape $x_s$ for each SN $s$.  A simple linear regression model for the absolute magnitude $M_s$ as a function of the distance-independent light curve observables is constructed using the Tripp formula\footnote{This formula is often written with an arbitrary negative sign preceding $\alpha$ \citep[e.g.][]{guy05, astier06}.  When it is regarded as a linear regression model for absolute magnitude versus the covariates $x$ and $c$, it is most natural to precede all the regression coefficients by a positive sign.}:
\begin{equation}\label{eqn:tripp1}
M_s = m_s - \mu_s = M_0 + \alpha \times x_s + \beta \times c_s + \epsilon_\text{res}^s
\end{equation}
\citep{tripp98}, where $\mu_s$ is the supernova distance modulus.  The global coefficients $(M_0, \alpha, \beta)$ are found by fitting this relation with measurements $(\hat{m}_s, \hat{c}_s, \hat{x}_s)$ on the Hubble diagram, $\mu_s = \mu(z_s)$, for a sample of SN Ia $\{s\}$.   The optical ``broader-brighter'' width-luminosity relation \citep{phillips93} is captured by $\alpha$, the ``redder-dimmer'' color-luminosity relation by $\beta$.  The expected absolute magnitude at $x_s = c_s = 0$ is $M_0$.  This equation would be correct if the color-luminosity relation were entirely due to (small amounts) of dust (and the light curve shape dependence truly linear).  In that case $m_s - \beta c_s$ is the reddening-free Wesenheit magnitude \citep{madore1982}.  However, in the presence of intrinsic color-luminosity variations, this formula is not fundamental: it is just the simplest linear model for absolute magnitude as a function of the observables, light curve shape and apparent color.   

The residual scatter\footnote{\label{footnote1}This variance term has been variously called the \emph{intrinsic} scatter or dispersion $\sigma_\text{int}$ \citep[e.g.][]{astier06, conley11, marriner11}.   However, in their conventional usage, there is no implication that this scatter is solely attributed to physical properties intrinsic to the SN Ia with host galaxy dust subtracted.  \citet{scolnic14} instead refers to it as \emph{residual} scatter: it is the additional variance needed to account for the scatter in the Hubble \emph{residuals}.  We adopt the latter usage to avoid confusion, and reserve \emph{intrinsic} within our model to conceptually refer to the latent properties of the SN Ia in the absence of host galaxy dust.} around this model that is unaccounted for by measurement error or peculiar velocities is $\epsilon_\text{res}^s$ with variance $\sigma^2_\text{res}$.  This contributes to the uncertainty in absolute magnitude $M_s$, and thus the photometric distance modulus $\mu_s$, of an individual SN.

In typical usage, $M_s$ and $m_s$ are peak absolute and apparent magnitudes effectively in rest-frame $B$-band, and the color corresponds to peak apparent $B-V$.  Hence $\beta$ is the slope of the change in $B$-magnitude for a unit change in $B-V$.   This can be compared to the expected slope for normal Milky Way dust extinction $A_B$ vs. $E(B-V)$ reddening, $R_B \equiv R_V + 1 = 4.1$.   

\citet{tripp98} and \citet{trippbranch99} originally found $\beta \approx 2$ using peak apparent $B-V$ colors and  $\Delta m_{15}(B)$ for light curve shape \citep{phillips93}.   Using the first SALT model to determine optical light curve stretch and color, \citet{guy05} and \citet{astier06} also found low values $\beta \approx 1.5-2$.  \citet{conley07} fit nearby SN Ia on the Hubble diagram and found that the empirical relation between SN Ia optical luminosity and apparent color, controlling for light curve shape, still required a low value of $\beta \approx 2$.  They speculated that this is much less than the normal dust $R_B \approx 4$ either because the dust in SN Ia hosts is nonstandard, or because the estimated $\beta$ may actually be measuring some combination of intrinsic color variations (not accounted for by light curve shape $x_s$) and normal interstellar dust.  

The SALT2 spectral template \citep{guy07,guy10} is the most popular \citep[and well-tested,][]{mosher14} model currently applied to fit cosmological SN Ia light curve data.   Using SALT2 and the Tripp formula to fit SNLS1 and SDSS-II SN Ia data, \citet{guy07} and \citet{kessler09} found $\beta \approx 1.8-2.6$.  For SNLS3, \citet{guy10} and \citet{conley11} found $\beta \approx 3.1$.  Similarly, the combined SDSS+SNLS3 [JLA] analysis of \citet{betoule14} obtained $\beta = 3.10 \pm 0.08$, significantly less than $R_B = 4.1$.

\subsection{Luminosity vs. Color Residual Scatter}

\citet{marriner11} introduced a more general formalism (\textsc{SALT2mu}) for accounting for the residual scatter around the linear model, Eq.\ref{eqn:tripp1}, and fitting for its regression coefficients.   Sensible methods for fitting Eq. \ref{eqn:tripp1} take into account the fact that the fitted values $(\hat{m}_s, \hat{c}_s, \hat{x}_s)$ are different from the true, latent (unobserved) values $(m_s, c_s, x_s)$ that obey Eq. \ref{eqn:tripp1}.  The difference amounts to random measurement error with the SALT2 fit covariance matrix.  \citet{marriner11} further supposes an additional source of residual scatter between the measured values and the latent values.  For example, the measured color of a SN $s$ is decomposed as 
\begin{equation}\label{eqn:salt2mu}
\hat{c}_s = c_\text{mod}^s + c_r^s + c_n^s,
\end{equation}
where $c_\text{mod}^s$ is the ``model'' color component that enters into the Tripp \emph{model} (Eq. \ref{eqn:tripp1}), and is linearly correlated with the luminosity, $c_n^s$ is the color measurement error ``noise'' and $c_r^s$ is a random ``residual'' color scatter term \citep{scolnic14}.   Similar equations can be written for the scatter in the magnitude and light curve shape components of the data.  While the measurement errors are quantified by a covariance matrix estimated from the SALT2 light curve fit, the residual scatter covariance matrix $\bm{\Sigma}_r$ is a priori unknown and poorly constrained by the data, and some choices must be made regarding its entries.  To date, it has been used generally with non-zero entries only for residual magnitude $\sigma_{m_r}^2$ and color $\sigma_{c_r}^2$ variances.   The special case in which only the residual magnitude variance entry is non-zero corresponds to the conventional assumption that the residual scatter is attributed to unexplained variance in luminosity, $\sigma_\text{res}^2 = \sigma_{m_r}^2$.  With an assumed residual matrix, \textsc{SALT2mu} estimates the $M_0$, $\alpha$, $\beta$ coefficients by minimizing the Hubble residual $\chi^2$, modified to incorporate measurement errors and the residual scatter matrix. 
Applying \textsc{SALT2mu} to SDSS-II data, \citet{marriner11} find that attributing the residual scatter only to color led to a larger estimated $\beta \approx 3.2$, still significantly less than $R_B = 4.1$.

\citet{scolnic14} analyzed a combined dataset, consisting of SDSS-II, SNLS3 and nearby samples, to examine the dependence of the estimated $\beta$ on the relative attribution of residual scatter to luminosity or color when analyzing the data within the \textsc{SALT2mu} framework.  When ``luminosity variation'' is assumed ($\sigma_{m_r} \gg \sigma_{c_r}$), the estimated $\beta \approx 3.2$ but when ``color variation'' is assumed ($\sigma_{m_r} \ll \sigma_{c_r}$), it increases to $\beta \approx 3.7$, closer to the normal MW average.    Simulations from a color variation model, in which $c_\text{mod}$ had a ``dust-like'' distribution (with a ``one-sided'' tail to the red, but a sharp edge to the blue), with a MW dust-like true $\beta = 4.1$, could better match the pattern of Hubble residuals vs. color seen in the data, compared to a luminosity variation simulation with a conventional $\beta = 3.1$.  They estimated that a misattribution of the residual scatter to luminosity rather than color variation, could lead to a bias $\Delta \beta \approx -1$ in the recovered slope, and a $4\%$ shift in the inferred $w$.  

Recently, \citet{scolnic16} presented a method to determine the underlying distribution of the latent $c_\text{mod}$ colors by matching the observed data distribution to realistic forward simulations that incorporate measurement noise, and selection effects, and two SALT2 spectral variation models: luminosity-variation dominated \citep[G10,][]{guy10} and color-variation dominated \citep[C11, based on][]{chotard11}.   Applying this to the cosmological SN Ia compilation of \citet{scolnic15}, they uncover a ``dust-like'' underlying color distribution, with a red tail and blue edge, when simulating with a color-dominated C11 model.   By matching their simulations to data, they find $\beta = 3.85$ using the color-variation model, and $\beta = 3.1$ with the luminosity-variation model.

These analyses suggest that the proper modeling of color-luminosity subcomponents is important for understanding the observed SN Ia color-magnitude distribution, the accurate estimation of distances, and inferences for cosmology.

\subsection{Shortcomings of these approaches}

The simplicity of the Tripp formula has enabled its widespread application in cosmological SN Ia analyses.   However, in its conventional usage, Eq. \ref{eqn:tripp1} is too simplistic.  Because neither the magnitude $m_s$ nor the color $c_s$ that enter into it are corrected for host galaxy dust extinction or reddening, the absolute magnitude $M_s = m_s - \mu_s$ is actually the dust-extinguished absolute magnitude $M^\text{ext}_s$, and $c_s$ is the dust-reddened apparent color $c^\text{app}_s$.  However, in reality, the extinguished absolute magnitude results from the dimming of the supernova's intrinsic luminosity by dust extinction $M^\text{ext}_s = M^\text{int}_s + A_B^s$.  The apparent color results from the dust reddening the supernova's intrinsic color: $c^\text{app}_s = c^\text{int}_s + E(B-V)_s$.  The dust extinction is surely correlated with its reddening, and can only be positive.  The supernova's intrinsic luminosity may be correlated with its intrinsic color, independently of light curve shape, as speculated by e.g. \citet{conley07} and \citet{freedman09}; at the very least we do not know that this intrinsic correlation is zero, and would like to estimate it.  By regressing only the sum $M^\text{ext}_s$ against the sum $c^\text{app}_s$, the Tripp formula tries to capture all the color-magnitude correlations in a single trend with slope $\beta$.   As it is a priori highly unlikely that the intrinsic color-magnitude slope would be exactly equal to the dust reddening-extinction law, this parameterization is clearly limited and inadequate. It fails to distinguish between the different physical characteristics of color-luminosity variation intrinsic to the SN Ia vs. reddening-extinction by extrinsic host galaxy dust.    

The residual matrix framework of \textsc{SALT2mu} adds some additional degrees of freedom to the conventional Tripp formula.  By decomposing the apparent color $c_s$ into a model color $c_\text{mod}$ and residual color $c_\text{r}$, and allowing $\sigma_{c_r} > 0$, one can in effect capture additional color-magnitude variations.  However, this residual color scatter  does not correlate with luminosity; this additional component essentially has its own $\beta_r = 0$.  Hence, the residual color scatter by itself would not be able to capture a separate non-zero color-luminosity correlation different from $\beta$.   This suggests that the color-magnitude covariance entry in the residual matrix $\bm{\Sigma}_r$ could be made non-zero.   This raises two challenges: either a numerical value for this residual covariance would have to be set a priori, or it would have to be inferred jointly with the other parameters.  Unfortunately, in the former case, it is unclear what value to set it to a priori; in the latter case, the inference of this covariance would likely be highly degenerate with $\beta$, unless strong priors were set.

Systematic uncertainties in the treatment of dust and color of SN Ia Ia have important implications for cosmological inference.  If the single slope $\beta$ is actually measuring a combination of intrinsic color-luminosity variation and host galaxy dust reddening-extinction, then different SN Ia subsamples may have different proportions of each.  The proportions may even be redshift-dependent, owing to the physical environment of the progenitor systems, host galaxies, or selection effects.  Applying one $\beta$ slope across the entire sample would incur complex systematic biases that propagate into cosmological inferences.   Resolving the confusion between the intrinsic variation and extrinsic host galaxy dust effects is imperative for the proper analysis of SN Ia observables.

\subsection{Simple-BayeSN}

To remedy the aforementioned shortcomings of the conventional methodology, we propose a new statistical model, \textsc{Simple-BayeSN}, describing the observed color-magnitude distribution as arising from the probabilistic combination of an intrinsic SN Ia color-magnitude variations and host galaxy dust reddening and extinction.  The host galaxy dust is given a physically-motivated distribution, allowing for only positive extinction.  The intrinsic color-magnitude slope $\beta_\text{int}$ can be different from the reddening-extinction slope $R_B$.  The observed data arises from the combination of these effects with measurement error.  By fitting this statistical model to the light curve data, the separate physical characteristics of the intrinsic and dust distributions are coherently inferred.  As our model uses the same SN Ia light curve measurements that are conventionally used by the Tripp formula, it can be readily applied to current cosmological SN Ia datasets.

We adopt a hierarchical Bayesian, or multi-level modeling, framework  to build a structured probability model conceptually describing the multiple random effects that underlie the observed SN Ia. This principled strategy enables us to coherently model and make probabilistic inferences at both the level of an ensemble or population of objects as well as at the level of the constituent individuals \citep{gelman_bda, loredohendry10, loredo12}.   Inference with the hierarchical model may be regarded as a probabilistic deconvolution of the observed SN data into the multiple, unobserved, latent random effects generating it \citep{mandel_scmav}.   Recent astrophysical and cosmological applications of hierarchical Bayesian modeling include \citet{foster13, brewer14, sanders14, mandel14, dfm14, schneider15, alsing16} and \citet{wolfgang16}.

Hierarchical Bayesian statistical modeling was first applied to SN Ia analysis by \citet{mandel09,mandel11}, who constructed the \textsc{BayeSN} model for optical and NIR SN Ia light curves \citep{mandelthesis}.   Fundamentally, \textsc{BayeSN} models the photometric time series observations of SN Ia as arising from intrinsic light curves and a host galaxy dust extinction across optical and NIR wavelengths.  The training process of \textsc{BayeSN} learns the covariance structure of  the intrinsic SN Ia light curve distribution across phase and wavelength, as well as the characteristics of the dust distribution and dust law $R_V$.  The latent variables for each SN, and the hyperparameters of the population distributions are inferred coherently from the joint posterior density conditional on the set of light curve data.  \citet{mandel11} demonstrated with \textsc{BayeSN} that the combination of optical and NIR observations could significantly improve constraints on SN Ia dust and the precision of photometric distances.

The \textsc{Simple-BayeSN} approach distills the core concept of \textsc{BayeSN}: hierarchically modeling the SN Ia data as a combination of intrinsic SN Ia variations, dust effects, and measurement error.  However, \textsc{BayeSN} is a complex framework that directly and non-parametrically models the multi-wavelength photometric time series observations.  As a simplification, \textsc{Simple-BayeSN} instead employs the outputs from an external light curve model that fits the photometric time series data of individual SN Ia and estimates the three parameters used in conventional analyses: peak apparent magnitude, apparent color, and light curve shape.  In this paper, we employ the widely-used SALT2 light curve model, but \textsc{Simple-BayeSN} can generally work with the fit parameters from any external model for apparent SN Ia light curve data.

\citet{march11} constructed the first hierarchical Bayesian model for fitting the cosmological SN Ia Hubble diagram with the SALT2 optical light curve fit parameters.   They encapsulated the Tripp formula in a hierarchical linear regression with population distributions for light curve shape and apparent color.  The regression coefficients and the population hyperparameters are inferred simultaneously with cosmological parameters in the posterior distribution.  This concept was further extended recently by \citet[UNITY,][]{rubin15} and \citet[BAHAMAS,][]{shariff16}.   However, these Bayesian models inherit the same fundamental, conceptual limitations of the Tripp formula, because they are essentially still regressing extinguished absolute magnitudes directly against apparent light curve parameters, rather than modeling the constituent, and physically distinct, intrinsic and dust components underlying the data.

This paper is structured as follows. In \S \ref{sec:motivation}, we construct a probabilistic generative model for the distribution of extinguished absolute magnitudes and apparent colors as a convolution of the intrinsic SN Ia color-magnitude distribution and the host galaxy dust distribution.    
We describe the features and generic implications of this model. 
In \S \ref{sec:model}, we encapsulate this generative model in a hierarchical Bayesian framework, \textsc{Simple-BayeSN}, for analyzing SN Ia magnitudes, colors and light curve shape measurements.  We demonstrate inference of the parameters of this model from the SN Ia data via maximum likelihood and Gibbs sampling.   In \S \ref{sec:application} we apply this model to analyze a data set of SALT-II parameters for a sample of 248 nearby SN Ia  ($z < 0.10$).  
In \S \ref{sec:hostmass}, we demonstrate how the host galaxy stellar mass dependence \citep{pkelly10} can be included in this new framework. We discuss our results in \S \ref{sec:discussion} and conclude in \S \ref{sec:conclusion}.  Mathematical and computational details about our methods are described in Appendices \S \ref{sec:app:dustydistr}, \ref{sec:bayesian}, \ref{sec:algorithm}, \& \ref{sec:fit_tripp}.

\vfill\break
\section{Motivation: \\A Probabilistic Generative Model}\label{sec:motivation}

In this section, we illustrate the essential concepts underlying our statistical approach by constructing a probabilistic generative model for the SN Ia color-magnitude relation.   We simulate a SN Ia sample of $N_\text{SN} = 250$ SN Ia uniformly distributed between redshifts $z = 0.01$ and $0.10$, and assume a fiducial $\Lambda$CDM cosmology of $h = 0.72$, $\Omega_M = 0.27$, $\Omega_\Lambda = 0.73$ and $w=-1$.  We assume that the optical light curve data of each SN Ia is fit with a light curve model, which returns three useful measurements in the rest-frame of the SN: the peak $B$-band apparent magnitude $\hat{m}_s$, the peak apparent $B-V$ color $\hat{c}_s = \hat{c}^\text{app}_s$, and a light curve shape parameter $\hat{x}_s$, plus estimates of their fitting uncertainties.   (The ``hatted'' quantities indicate estimated or measured values, which differ from the true values by some random measurement error).   

To focus on understanding the effects of intrinsic color-magnitude variations and host galaxy dust, we will assume that the the observables,  the apparent magnitude, apparent color and light curve shape, for each SN $s$ are estimated without error, $(\hat{m}_s, \hat{c}^\text{app}_s, \hat{x}_s) = (m_s, c^\text{app}_s, x_s)$, and that peculiar velocities are negligible: $\sigma_\text{pec} = 0 \text{ km s}^{-1}$.  Hence, conditional on the redshift and cosmological parameters, the distance moduli are known, as are the extinguished absolute magnitudes: $M^\text{ext}_s = m_s - \mu(z_s)$.  In \S \ref{sec:model}, we will construct a statistical model for analyzing observed data accounting for measurement error and peculiar velocities, inference of parameters and prediction of photometric distances.

To simulate the data, we generate values of the latent variables of intrinsic color $c^\text{int}_s$, intrinsic absolute magnitude $M^\text{int}_s$, light curve shape $x_s$, and host galaxy dust reddening (or color excess) $E_s = E(B-V)$ for each supernova $s$.  In our usage, \emph{latent} variables or parameters are those physical quantities for each supernova that are unobserved, but underlie the observed measurements.   \emph{Intrinsic} parameters refer to the properties of the supernova in the absence of host galaxy dust effects and measurement error.   We assume simple forms for the population distributions underlying these latent variables.   These distributions are governed by a set of \textit{hyperparameters}, describing the intrinsic SN Ia population distribution and the host galaxy dust distribution.  Data or parameters pertaining to individual supernovae are denoted by a subscript $s$, whereas hyperparameters common to the population are not.  We adopt values that are similar to those estimated later in $\S \ref{sec:application}$ by applying our model to the observed data.

\subsection{Intrinsic Absolute Magnitudes and Colors}

\begin{figure}[t]
\centering
\includegraphics[angle=0,scale=0.38]{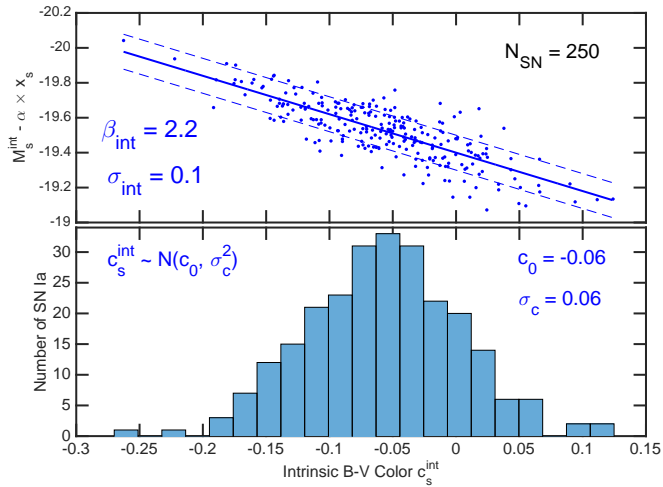}
\caption{\label{fig:fig1} Simulation of intrinsic supernova quantities. (bottom) Simulated intrinsic colors $\{c^\text{int}_s\}$ drawn from a Gaussian distribution.  (top) Distribution of $x_s$-corrected intrinsic absolute magnitudes and intrinsic colors $\{M^\text{int}_s, c^\text{int}_s\}$. The blue line corresponds to the intrinsic color-magnitude slope $\beta_\text{int}$ and the dashed lines correspond to the intrinsic scatter ($\pm \sigma_\text{int})$.}
\end{figure}

We generate light curve shape parameters $\{x_s\}$ for each SN, drawn from a Gaussian population distribution with mean $x_0 = -0.40$ and variance $\sigma_x^2 = (1.2)^2$.   A set of intrinsic colors ($B-V$ at maximum light) $\{c^\text{int}_s\}$ is drawn from an assumed Gaussian population distribution with mean $c_0^\text{int} = -0.06$ mag and variance $\sigma_{c,\text{int}}^2 = (0.06 \text{ mag})^2$.   We assume that the intrinsic absolute magnitude (in rest-frame $B$-band), $M_s$, is simply related to these quantities by a linear relation:
\begin{equation}\label{eqn:Mint_sim}
M^\text{int}_s = M_0^\text{int} + \alpha x_s + \beta_\text{int} c^\text{int}_s + \epsilon^\text{int}_s
\end{equation}
where $M_0^\text{int}$ is the expected intrinsic absolute magnitude for an $(x_s = 0, c^\text{int}_s = 0)$ supernova, $\alpha$ is the slope of the Phillips light curve width-luminosity relation, and $\beta^\text{int}$ is the slope against intrinsic color.  We adopt true values of $M_0^\text{int} = -19.40$ mag, $\alpha = -0.15$ and $\beta^\text{int} = 2.2$.  The random scatter around the mean relation is $\epsilon^\text{int}_s \sim N(0, \sigma_\text{int}^2)$ with variance $\sigma_\text{int}^2 = (0.10 \text{ mag})^2$.  In Figure \ref{fig:fig1}, we show the result of simulating these intrinsic supernova quantities from this generative model in this manner.  Since we are focusing on effects in the color-magnitude plane, we control for light curve shape by plotting the light curve shape-corrected intrinsic absolute magnitude $M^\text{int}_s - \alpha \times x_s$ versus intrinsic color $c^\text{int}_s$.

\subsection{Host Galaxy Dust Extinction and Reddening}

As the light of the supernova leaves its host galaxy, it passes along the line-of-sight through a random column density of interstellar dust, which absorbs and scatters the light in wavelength-dependent process, resulting in extinction and reddening.  The resulting change in rest-frame $B-V$ supernova color is the color excess or dust reddening and is denoted $E_s = E(B-V)$ for SN $s$.  The resulting dimming is quantified by the change in the peak absolute magnitude in $B$-band by the extinction $A_B$.  The dust extinction and color excess are related by the parameter $R_B$, a property of the dust: $A_B^s = R_B E_s$.

Since dust only dims and reddens, $E_s$ is a positive quantity.  We assume that the dust reddening for each SN Ia is randomly drawn from an exponential population distribution with population mean $\tau = 0.07$ mag: $E_s \sim \text{Expon}(\tau)$.
The exponential distribution for the dust population has been previously used by, e.g. \citet{jha07, mandel09, mandel11, mandel14}.  
The effect of dust is to dim the absolute magnitude
\begin{equation}\label{eqn:Mext_sim}
M^\text{ext}_s = M^\text{int}_s + R_B E_s,
\end{equation}
and redden the color
\begin{equation}\label{eqn:capp_sim}
c^\text{app}_s = c^\text{int}_s + E_s,
\end{equation}
of SN $s$.   We assume for simplicity that the same $R_B$ value characterizes the dust in all SN Ia host galaxies.  

\begin{figure}[t]
\centering
\includegraphics[angle=0,scale=0.425]{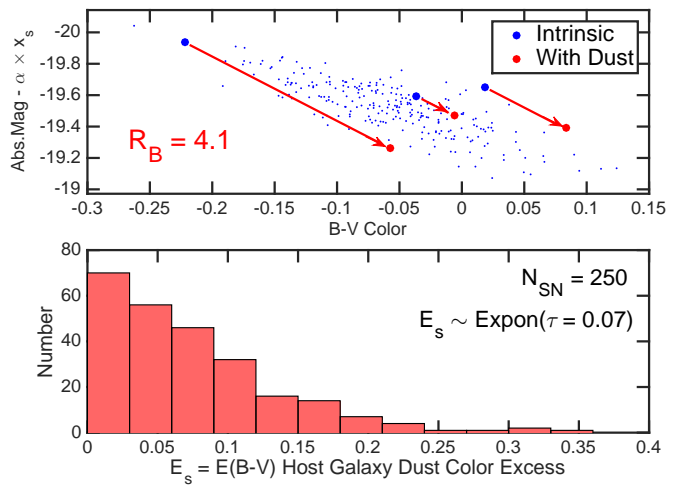}
\caption{\label{fig:fig2} Simulation:   (bottom) Distribution of dust reddening $\{E_s\}$, simulated from an exponential distribution with mean $\tau = 0.07$ mag. (top) The effect of host galaxy dust reddening and extinction on the intrinsic SN Ia magnitudes and colors.  The intrinsic (blue) points are the same as those in Fig. \ref{fig:fig1}.  Each intrinsic (blue) point maps to a extinguished magnitude-apparent color (red) point through the extinction and reddening effects of dust.  The slope of the red extinction-reddening vectors is $R_B = 4.1$. The length of each red vector is given by a random value of the reddening $E_s$, drawn from the exponential distribution.  We illustrate this for three random SN Ia (large points).}
\end{figure}

\begin{figure}[b]
\centering
\includegraphics[angle=0,scale=0.4]{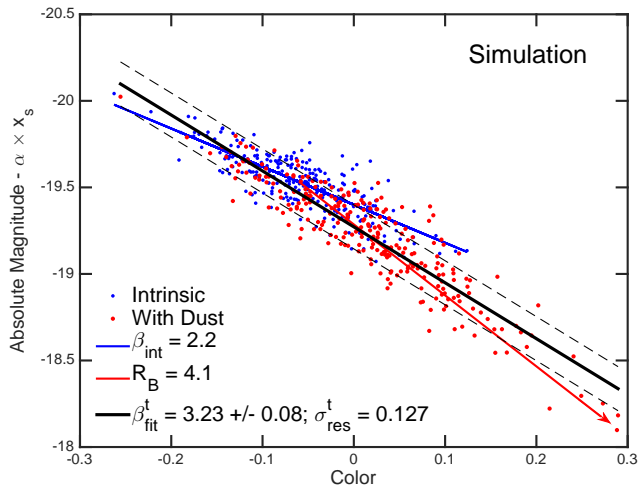}
\caption{\label{fig:fig3} Simulation: The intrinsic SN Ia magnitude-color distribution (blue points) is the same as that in Figs. \ref{fig:fig1} and \ref{fig:fig2}, and has an intrinsic slope $\beta_\text{int} = 2.2$.  The effect of host galaxy dust drawn from an exponential distribution $E_s \sim \text{Expon}(\tau)$ is to map each blue point (intrinsic magnitude and intrinsic color) to a red point (extinguished magnitude and apparent color) along the dust vector (red arrow) with slope $R_B = 4.1$. When the conventional linear Tripp model is used to fit the relation between extinguished magnitudes and apparent colors, the resulting slope (black line) is $\beta^t = 3.23 \pm 0.08$ with residual scatter $\sigma^t_\text{res} = 0.127$ mag.  The result of the conventional linear fit returns a value that is neither the true intrinsic slope, nor the dust slope, but a weighted average between the two.}
\end{figure}

In Figure \ref{fig:fig2}, we show the distribution of host galaxy dust reddening $E_s$ drawn from assumed exponential distribution.  We adopt a value of $R_B = 4.1$ and demonstrate the effect of reddening and extinction on the intrinsic magnitudes and color.  Each intrinsic (blue) point maps to a red point through the effects of dust. 
In Figure \ref{fig:fig3}, we show the resulting distribution of extinguished absolute magnitudes (controlling for light curve shape) $M^\text{ext}_s - \alpha \times x_s$ versus apparent color $c^\text{app}_s$ as red points, as well as the original intrinsic distribution (with mean slope $\beta_\text{int} = 2.2$.  The red arrow indicates the trajectory (with slope $R_B = 4.1$) of a SN with average intrinsic properties under the effect of increasing host galaxy dust.

\subsection{Implications of Inference with the Tripp Model}

Next, we analyze the data by fitting the linear Tripp formula, as is conventionally done:
\begin{equation}\label{eqn:tripp}
m_s - \mu_s = M^\text{ext}_s = M_0^{\text{ext},t} + \alpha^t x_s + \beta_\text{app}^t c^\text{app}_s +\epsilon_{\text{res},s}^t
\end{equation}
where the residual scatter about the relation is $\epsilon_{\text{res},s}^t \sim N(0,(\sigma^t_\text{res})^2)$.  To improve clarity, we introduce additional notation compared to Eq. \ref{eqn:tripp1}.   The absolute magnitude and color are actually the dust-extinguished absolute magnitude $M^\text{ext}_s$  and dust-reddened apparent color $c^\text{app}_s$.  The regression coefficients $(M_0^{\text{ext},t}, \alpha^t, \beta_\text{app}^t)$ and the residual variance $\sigma^t_\text{res}$ are labelled by superscripts$^t$ (for Tripp) to distinguish them from the hyperparameters of the generative model.   

We estimate $(M_0^t, \alpha^t, \beta^t, \sigma^t_\text{res})$ from the simulated data via maximum likelihood.  The resulting fit in the color-magnitude plane is shown as the black line, with slope $\beta^t = 3.23 \pm 0.08$.  The residual scatter around the black line is $\sigma^t_\text{res} = 0.127$ mag.  We find that when the true intrinsic magnitude-color slope is $\beta_\text{int} = 2.2$, and the true dust law slope is $R_B = 4.1$, the linear Tripp estimator obtains neither.  Rather, it fits a value somewhere in-between the true intrinsic slope and the dust law.   For different values of the true hyperparameters, the value that $\beta_\text{app}^t$ will obtain depends on the values of $R_B$ and $\beta_\text{int}$, as well as the intrinsic color dispersion $\sigma_{c,\text{int}}$, the intrinsic scatter $\sigma_\text{int}$ and the average amount of dust extinction $R_B \tau$ in the host galaxies.

In the very red (positive) tail of the apparent color distribution, there are more red points below the fitted Tripp relation than above.  This can also be seen in the blue (negative tail) of the apparent color distribution.  This indicates that the linear Tripp model gives a biased trend of the extinguished absolute magnitude with apparent color for supernovae in the tails of the apparent color distribution.  Furthermore, if the true mean trend of extinguished absolute magnitude vs. apparent color were actually linear, then the naive linear model would have captured this, resulting in no bias in the tails.  This suggests that the true mean trend of $M^\text{ext}$ vs. $c^\text{app}$, under the true generative model, must be non-linear as a function of $c^\text{app}$ (for a fixed light curve shape $x$).  

We can calculate this trend given the mathematical model for the generative process we have just described.   The extinguished absolute magnitudes and apparent colors result from adding dust extinction and reddening to the intrinsic absolute magnitudes and intrinsic colors.  Therefore, the ``dusty'' distribution of extinguished absolute magnitudes and apparent colors is a convolution of the intrinsic SN Ia distribution and the host galaxy dust distribution\footnote{If random variables $X$ and $Y$ are drawn from probability distributions $P_X$ and $P_Y$, respectively, then the probability distribution of their sum, $X+Y$ is the convolution of $P_X$ and $P_Y$.}.   From this convolved distribution, the mean trend of extinguished absolute magnitude with apparent color for a given light curve shape can be computed (see Appendix \S \ref{sec:app:dustydistr}).   The trend has the properties that in the blue limit ($c^\text{app} \rightarrow -\infty$) its derivative smoothly approaches the intrinsic slope $\beta_\text{int}$ and in the red limit ($c^\text{app} \rightarrow +\infty$) its slope approaches the dust law $R_B$.  The linear Tripp formula approximates an intermediate slope. 

\begin{figure}[t]
\centering
\includegraphics[angle=0,scale=0.43]{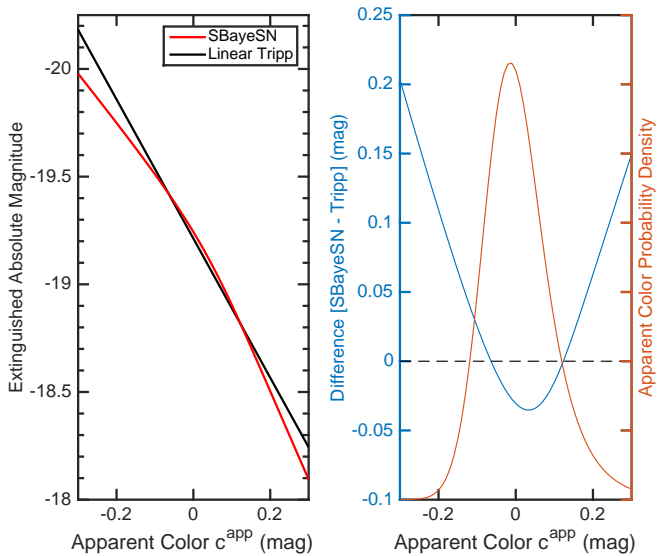}
\caption{\label{fig:fig4} Simulation: (left)  The true mean trend of extinguished absolute magnitude as a function of apparent color (for an average light curve shape $x_s = x_0$). The black line is the linear fit to the simulated data using the conventional Tripp formula, with slope $\beta^t = 3.23$.  The red line is the expected trend under the \textsc{Simple-BayeSN} generative model with $\beta_\text{int} = 2.2$ and $R_B = 4.1$.   This trend smoothly transitions between a slope of $\beta_\text{int}$ in the blue (negative) tail, and a slope of $R_B$ in the red (positive) tail.   The $\beta_{app}^t$ in the linear Tripp formula approximates the derivative near the middle of the apparent color distribution. (right)  The difference between the trend under the true model and the linear fit is shown as the solid blue curve.  
The apparent color distribution is shown as the orange curve.}
\end{figure}

In the left panel of Figure \ref{fig:fig4}, we plot the trend of $M^\text{ext}$ vs. $c^\text{app}$,  computed under our generative model (\textsc{Simple-BayeSN} or \textsc{SBayeSN}, red line), and the color-magnitude relation modeled by the conventional linear Tripp formula (black line) for an SN with an average light curve shape $x = x_0$, using the values fit from the simulation.   We see that the apparent color-magnitude slope $\beta^t_\text{app}$ obtained by the linear Tripp estimator approximates the derivative of the true curve near the middle of the apparent color distribution.  In the right panel, we show the difference between the trend under our model and the linear fit.  We also overplot the true apparent color population distribution, Eq. A5 of \citet{mandel14}, using the true values of ($c_0^\text{int}$, $\sigma_{c,\text{int}}$, $\tau$).

With respect to the nonlinear prediction of the true generative model, the naive linear model systematically \textit{overestimates} the extinguished absolute luminosities, and thus the distances, of the SN Ia with very blue (negative) or very red (positive) apparent colors.  It also slightly systematically \textit{underestimates} the extinguished absolute luminosities, and thus the distances, of the SN Ia with average apparent color.  Although these biases are small relative to the variance $\sigma^t_\text{res}$ for one supernova, they will \emph{not} decrease with increasing sample size. 

\subsection{Comparison to the Tripp Formula}\label{sec:comp_tripp}

In the absence of measurement error, we can substitute Eq. \ref{eqn:capp_sim} and Eq. \ref{eqn:Mext_sim} into Eq. \ref{eqn:Mint_sim} to obtain
\begin{equation}\label{eqn:m_alt}
m_s - \mu_s = M_0^\text{int} + \alpha x_s + \beta_\text{int} c^\text{app}_s + (R_B - \beta_\text{int}) E_s + \epsilon^\text{int}_s 
\end{equation}
By comparing this to Eq. \ref{eqn:tripp}, we see that if the dust law $R_B$ is equal to the intrinsic color-magnitude slope $\beta_\text{int}$, then our model reduces to the Tripp formula, up to relabelling of the fit parameters, since $(R_B - \beta_\text{int})E_s = 0$ for every value of $E_s$, regardless of the distribution of $E_s$.  In this case, intrinsic color effects are effectively indistinguishable from dust effects, at least within this model.  The regression coefficients should match $(M_0^\text{int}, \alpha, \beta_\text{int}) = (M_0^\text{ext,t}, \alpha^t, \beta^t_\text{app})$ when fit to the data.

If the intrinsic color-magnitude slope is differs from the dust law $R_B \neq \beta_\text{int}$, then the regression equation Eq. \ref{eqn:m_alt} is (up to relabelling of fit parameters) the Tripp formula \textbf{plus} an extra random variable $(R_B - \beta_\text{int})E_s$.  If $R_B > \beta_\text{int}$, as we find in this paper, then this quantity is always positive.  (At other wavelengths, it may be that $R < \beta_\text{int}$, in which case it would always be negative).  Since dust only reddens, $E_s > 0$, this term could be regarded as an extra ``noise'' term that has a non-zero mean and an asymmetric distribution (unlike $\epsilon^\text{int}_s$, which is assumed to be Gaussian and symmetric).  Neglecting this additional random noise term will result in biased estimates of the other parameters, so that generally $(M_0^\text{int}, \alpha, \beta_\text{int}) \neq (M_0^\text{ext,t}, \alpha^t, \beta^t_\text{app})$.

The simulation above demonstrates the inherently probabilistic nature underlying the apparent color-magnitude distribution. Consider, for example, a blue SN Ia with a well-measured light curve shape and apparent color $c_s^\text{app} = -0.10$ .  There is some probability that it is unaffected by host galaxy dust ($E_s = 0$) and its intrinsic color is also $c_s^\text{int} = -0.10$.  But there is also some probability that its intrinsic color is actually $c_s^\text{int} = -0.20$, but it suffers from $E(B-V)_s = +0.10$ mag of host galaxy reddening.   Since the intrinsic slope and the dust law are different, $\beta_\text{int} \neq R_B$, these two scenarios \emph{should} result in different calculations for $M^\text{ext}_s$, which \emph{should} in turn yield different distances $\mu_s$, given a well-measured apparent magnitude $m_s$.  

The Tripp formula, and indeed, any statistical model directly modeling a functional form of extinguished magnitude vs. apparent color and light curve shape,
\begin{equation}
m_s - \mu_s = M_s^\text{ext} = f(c_s^\text{app}, x_s),
\end{equation}
would predict the same distance in either of these two scenarios.  What is needed is a statistical model that properly weighs these different possibilities by their probabilities and marginalizes over them to produce a probability distribution for the photometric distance.

\vfill\break
\section{Simple-BayeSN: A Simple Hierarchical Bayesian Model for SN Ia}\label{sec:model}

In the previous section, we made some simple assumptions about the underlying probabilistic processes generating the SN Ia data, including the possibility that the intrinsic SN Ia magnitude-color slope $\beta_\text{int}$ is different from the dust law, $R_B$.  To test whether the data is consistent with two different slopes, we need to estimate the parameters governing the underlying processes.  To do this, we first construct a hierarchical Bayesian statistical model to describe the distribution of the observed data as a probabilistic convolution of the the intrinsic variations and dust effects.  We also include random effects such as measurement error and peculiar velocity uncertainties.  We derive the likelihood function and joint posterior probability of the latent variables and hyperparameters.  The hierarchical model is fit to the observed data with this model to estimate the hyperparameters using maximum likelihood and Gibbs sampling.

We regard as the observed data the peak apparent magnitude $\hat{m}_s$, the peak apparent color $\hat{c}^\text{app}_s$, and the light curve shape $\hat{x}_s$ for each SN $s$, obtained from fitting the SN Ia light curve (time series) data, as well as the measured redshift $z_s$.  The latent parameters for each SN $s$ are the observable light curve parameters $\bm{\phi}_s = (m_s, c^\text{app}_s, x_s)$, the dust reddening $E_s$ and the distance modulus $\mu_s$.   The hyperparameters of the dust distribution are the population mean dust reddening and the dust law parameter $\bm{\Theta}_\text{dust} = (\tau, R_B)$.  The hyperparameters $\bm{\Theta}_\text{SN}$  governing the intrinsic SN Ia correlations and distributions are described in \S \ref{sec:intrinsic_model}.  

\subsection{Light Curve Fitting Error Likelihood}

The observed data for conventional SN Ia analysis consists of broadband optical SN Ia light curves or brightness time series.  A statistical model for the SN Ia time series (``light curve fitter'') is fit to these data and returns estimates $\bm{d}_s$ of parameters useful for summarizing the light curve, as well as an estimate of their joint estimation uncertainty, $\bm{W}_s$, for each SN $s$.  The light curve fitter accounts for the redshifting of the SN Ia spectral energy distribution into the observed photometric passbands, and dust extinction from our Milky Way Galaxy.  We refer to these light curve parameter estimates as the light curve ``data,'' rather than the original multi-filter time series observations, and consider the error in these parameter estimates from the light curve fit as the ``measurement error.''

Our likelihood function for light curve fitting and estimation models the probability of the data $\bm{d}_s = (\hat{m}_s, \hat{c}^\text{app}_s, \hat{x}_s)^T$ given the latent observable parameters $\bm{\phi}_s$ as Gaussian with covariance matrix $\bm{W}_s$.
\begin{equation}\label{eqn:lc_lkhd}
P( \bm{d}_s | \, \bm{\phi}_s ) = N( \bm{d}_s |  \bm{\phi}_s, \, \bm{W}_s).
\end{equation}
For the analysis in this paper, we use SN Ia light curve parameter estimates and error covariance matrices  obtained from the SALT2 \citep{guy10} light curve fitter.  However, our hierarchical Bayesian statistical model can be used with parameter outputs from any suitable light curve fitter with the above characteristics.

\vfill\break
\subsection{Redshift-Distance Likelihood}

The expected theoretical distance modulus at redshift $z$ in a smooth cosmology with parameters $\bm{\Omega} = (h, \Omega_M, \Omega_\Lambda, w)$ is $\mu_{\Lambda CDM}(z; \bm{\Omega}) = 25 + 5 \log_{10}[ d_L(z; \bm{\Omega}) \text{ Mpc}^{-1}]$, where $d_L(z; \bm{\Omega})$ is the theoretical luminosity distance.  The Hubble Constant is $H_0 = 100h \text{ km s}^{-1} \text{ Mpc }^{-1}$.  If $z_s$ is the measured redshift of the SN host galaxy, the redshift-distance likelihood function is
\begin{equation}
\begin{split}
P(z_s | \, \mu_s, \bm{\Omega}) &= N[ z_s | \, \mu_{\Lambda CDM}^{-1}(\mu_s; \bm{\Omega}), \sigma_\text{pec}^2/c^2 + \sigma_z^2] \\
&\approx N[ \mu_{\Lambda CDM}(z_s; \bm{\Omega}) | \,  \mu_s, \sigma_{\mu|z,s}^2] .
\end{split}
\end{equation}
 The approximation was made in linearizing the distance modulus at the cosmological redshift $f(z^c_s; \bm{\Omega})$ about the observed redshift $z_s$.
The uncertainty in $\mu$ given the redshift is significant for low-$z$ objects, for which
\begin{equation}
\sigma_{\mu|z,s}^2 \approx  [5/ (z_s \ln 10)]^2 [\sigma_z^2 + \sigma_\text{pec}^2 / c^2]
\end{equation}
where $\sigma_z$ is the redshift measurement error and $\sigma_\text{pec} = 200 \text{ km s}^{-1}$ is the peculiar velocity dispersion \citep{carrick15}.  With $P(\mu) \propto 1$, this leads to 
\begin{equation}\label{eqn:zmu_lkhd}
P(\mu_s | z_s, \bm{\Omega})  = N[ \mu_s | \, \mu_{\Lambda CDM}(z_s; \bm{\Omega}), \sigma_{\mu|z,s}^2 ].
\end{equation}
For the analysis of nearby SN Ia in this study, we fixed  $\bm{\Omega} = \bm{\hat{\Omega}} = (0.72, 0.27,0.73,-1)$ to its concordance values.  It is important to remember that, because SN Ia by themselves are only \emph{relative} distance indicators, absolute magnitudes and distance moduli are only determined up to an overall additive constant.  Thus, a change in $h$ will trivially shift all the absolute magnitudes and distances accordingly.  However, a correct marginalization (either numerical or analytical) over the absolute magnitude constant $M_0$ (or the quantity $\mathcal{M}_0 = M_0 - 5 \log_{10} h$) removes the dependence on $h$ from all other inferences from the SN Ia data.

\subsection{Latent Variable Equations}

Define an \emph{intrinsic parameters} vector $\bm{\psi}_s =  (M^\text{int}_s, c^\text{int}_s, x_s)^T$, consisting of the intrinsic absolute magnitude, intrinsic color and light curve shape parameter.  These are related to the observable parameters through the effects of distance and host galaxy dust: $m_s = M^\text{ext}_s  + \mu_s = M^\text{int}_s + R_B E_s + \mu_s$ and $c^\text{app}_s = c^\text{int}_s + E_s$.  These can be written as a vector equation:
\begin{equation}
 \left(
\begin{array}{c}
m_s \\ c^\text{app}_s \\ x_s
\end{array}     \right) 
= \left( \begin{array}{c}
M^\text{int}_s \\ c^\text{int}_s \\ x_s
\end{array}     \right) +
\left( \begin{array}{c}
1 \\ 0 \\ 0
\end{array}     \right) \mu_s +
\left( \begin{array}{c}
R_B \\ 1 \\ 0
\end{array}     \right) E_s
\end{equation}
%
Defining the \emph{observable} parameter vector as $\bm{\phi}_s = (m_s, c^\text{app}_s, x_s)^T$, we can write this concisely as:
\begin{equation}
\bm{\phi}_s = \bm{\psi}_s + \bm{e}_1 \mu_s + \bm{e}_E E_s
\end{equation}
where $\bm{e}_E \equiv \bm{e}_2 + R_B \bm{e}_1$ and $\bm{e}_i$ is a unit vector along the $i$th coordinate axis.

It is not exactly true that the effect of applying a dust reddening law with a given extinction $A_V$ to SN Ia spectra will have linear effect on the magnitudes and colors measured through broadband filters.  However, linearity is a good approximation up to moderate extinction values \citep*{jha07}.

\subsection{Intrinsic SN Ia  Population Distribution}\label{sec:intrinsic_model}

We construct a simple model for the joint population distribution of the intrinsic supernova parameters, $P( \bm{\psi}_s | \, \bm{\Theta}_\text{SN})$, depending on some hyperparameters $\bm{\Theta}_\text{SN}$.  It is convenient to express this joint distribution in terms of a product of conditional probability densities, each of which we model separately.  
\begin{equation}\label{eqn:intrinsic_factorization}
\begin{split}
P( \bm{\psi}_s | \, \bm{\Theta}_\text{SN}) &=P( M^\text{int}_s, c^\text{int}_s, x_s| \, \bm{\Theta}_\text{SN}) \\
&= P(M^\text{int}_s | \, c^\text{int}_s, x_s; \bm{\Theta}_\text{SN}) \\
&\times P( c^\text{int}_s | x_s; \bm{\Theta}_\text{SN}) \\
&\times P( x_s | \, \bm{\Theta}_\text{SN})
\end{split}
\end{equation}
We can model each conditional factor with a mean relation and (Gaussian) scatter about the relation:
\begin{equation}
P(M^\text{int}_s | \, c^\text{int}_s, x_s; \bm{\Theta}_\text{SN}) = N[ M^\text{int}_s  | \, f_M(c^\text{int}_s, x_s; \bm{\Theta}_\text{SN}), \sigma_\text{int}^2]
\end{equation}
\begin{equation}
P(c^\text{int}_s |\, x_s; \bm{\Theta}_\text{SN}) = N[ c^\text{int}_s | \,f_c( x_s; \bm{\Theta}_\text{SN}), \sigma_{c,\text{int}}^2]
\end{equation}
\begin{equation}
P( x_s | \, \bm{\Theta}_\text{SN}) = N( x_s |\, x_0, \sigma_x^2).
\end{equation}
and the marginal population distribution of the light curve shapes $x$ as a Gaussian with mean $x_0$ and variance $\sigma_x^2.$  We adopt the most general linear model for the mean intrinsic absolute magnitude trend $f_M(c^\text{int}_s, x_s; \bm{\Theta}_\text{SN})$, and the mean intrinsic color trend $f_c(x_s; \bm{\Theta}_\text{SN})$.  Hence, the model for the intrinsic SN Ia parameters can be written as a hierarchy of linear equations.
\begin{equation}\label{eqn:intrinsic1}
M^\text{int}_s = M_0^\text{int} + \alpha x_s + \beta_\text{int} c^\text{int}_s + \epsilon_s^\text{int}
\end{equation}
\begin{equation}\label{eqn:intrinsic2}
c^\text{int}_s = c_0^\text{int} + \alpha_c^\text{int} x_s + \epsilon_s^{c,\text{int}}
\end{equation}
\begin{equation}\label{eqn:intrinsic3}
x_s = x_0 + \epsilon_s^x
\end{equation}
where the intrinsic scatter terms are $\epsilon_s^x \sim N(0, \sigma_x^2)$, $\epsilon_s^{c,\text{int}} \sim N(0, \sigma_{c,\text{int}}^2)$, $\epsilon_s^\text{int} \sim N(0, \sigma_\text{int}^2)$.
Equation \ref{eqn:intrinsic1} is similar to the linear Tripp formula but relates the intrinsic absolute magnitude to the light curve shape and intrinsic color.

To summarize, the nine hyperparameters governing the structure of the population distribution for the intrinsic SN Ia parameters are 
\begin{equation}
\bm{\Theta}_\text{SN} = (M_0^\text{int}, \alpha, \beta_\text{int}, \sigma_\text{int}^2, c_0^\text{int}, \alpha_c^\text{int}, \sigma_{c,\text{int}}^2, x_0, \sigma_x^2)
\end{equation}

\begin{itemize}
\item $M_0^\text{int}$ : the intrinsic absolute magnitude constant is the expected intrinsic absolute magnitude for a SN with light curve shape $x_s = 0$ and intrinsic color $c_s^\text{int} = 0$,
\item $\alpha$ : the slope of the trend of intrinsic absolute magnitude vs. light curve shape,
\item $\beta_\text{int}$ :  the slope of the trend of intrinsic absolute magnitude vs. intrinsic color, 
\item $\sigma_\text{int}^2$ : the intrinsic variance around the mean trend of intrinsic absolute magnitude vs. light curve shape and color,
\item $c_0^\text{int}$ : the expected intrinsic color for a SN Ia with light curve shape $x = 0$.  If $\alpha_c^\text{int} = 0$, then $c_0^\text{int}$ is the population mean intrinsic color,
\item $\alpha^\text{int}_c$ : the slope of the trend of intrinsic color vs. light curve shape,
\item $\sigma_{c,\text{int}}^2$ : the intrinsic variance around the mean trend of intrinsic color vs. light curve shape,
\item $x_0$ : the mean of the $x$ light curve shape population distribution,
\item $\sigma_x^2$ : the variance of the $x$ light curve shape population distribution. 
\end{itemize}

These model choices can be expanded upon by including additional variables such as spectroscopic indicators or host galaxy properties, or non-linear trends in modeling the population distribution Eq. \ref{eqn:intrinsic_factorization}.  For example, in \S \ref{sec:hostmass}, we modify Eq. \ref{eqn:intrinsic1} so that the intrinsic absolute magnitude constant depends on host galaxy stellar mass.

\subsection{Host Galaxy Dust Population Distribution}\label{sec:dust_model}

The host galaxy dust reddening $E_s \equiv E(B-V)$ is assumed to be drawn from an exponential population distribution with average $\tau$: $E_s \sim \text{Expon}(\tau)$.  This has a probability density only on positive redding $E_s > 0$ because dust only causes dimming and reddening:
\begin{equation}
P(E_s | \tau)  = \begin{cases} \tau^{-1} \exp(E_s / \tau), & E_s \ge 0 \\ 0, & E_s < 0 \end{cases}
\end{equation}
\citet{mandel11} found that this model describes well the distribution of peak apparent $B-V$ colors of nearby SN Ia up to $A_V < 1$.  Futhermore, we assume that the properties of the dust in host galaxies are described by the ratio of extinction to reddening $R_B = A_B / E(B-V)$.  The dust hyperparameters are $\bm{\Theta}_\text{dust} = (\tau, R_B)$:
\begin{itemize}
\item $\tau$ : The population average of the exponential distribution of dust reddening: $\tau = \langle E_s \rangle$. 
\item $R_B$: The ratio of $A_B$ dust extinction to $E(B-V)$ reddening.
\end{itemize}
These simple choices can be expanded upon in the future to allow for the dust distribution or $R_B$ to vary within subpopulations.  For example, in \S \ref{sec:hostmass}, we allow the average reddening $\tau$ to depend on host galaxy stellar mass.

\subsection{Hyperpriors}

We need to specify the prior in the hyperparameters, or hyperprior $P(\bm{\Theta})$. For $\theta_i = \sigma_\text{int}^2$, $\sigma_{c,\text{int}}^2$, $\sigma_x^2$, or $\tau$, we use flat priors on positive values only, $\theta_i \sim U(0,\infty)$.  For all other hyperparameters we use flat priors on positive and negative values, $\theta_i \sim U(-\infty,\infty)$.  In our application, our posterior inferences are insensitive to using proper uniform hyperpriors over a wide range.

\subsection{Bayesian Inference and Parameter Estimation}

In Appendix \S \ref{sec:bayesian}, we use the assumptions of the model to derive the marginal likelihoods and posterior distribution of all the parameters and hyperparameters given the light curve data and redshifts.   The conceptual relationships between the hyperparameters, latent parameters and data are depicted as a probabilistic graphical model in Fig. \ref{fig:fig16}.

We use two methods to estimate the hyperparameters (``train'' the model) from the light curve data $\mathcal{D} = \{ \bm{d}_s \}$ and redshifts $\mathcal{Z} = \{ z_s \}$ for a fixed cosmological model $\bm{\Omega} = \bm{\hat{\Omega}}$.   The first method finds the peak of the log marginal likelihood $P( \bm{d}_s | \, z_s; \bm{\Theta}, \bm{\hat{\Omega}} )$ in the 11-dimensional parameter space $\bm{\Theta}$ given by Eq. \ref{eqn:marglkhd} using a constrained nonlinear optimization algorithm.   The constraints are that $\sigma_x, \sigma_c, \sigma_\text{int},  \tau > 0$.  The uncertainties in the maximum likelihood values $\bm{\Theta}_\text{MLE}$ are estimated from the inverse Hessian of the negative log likelihood (the Fisher information matrix), although these are only asymptotically lower bounds on the parameter uncertainties.

To explore the parameter space and obtain a more complete measure of the joint parameter uncertainty, we run a Gibbs sampler to generate a Markov chain that converges to the global posterior probability Eq. \ref{eqn:globalposterior}.  This generates an MCMC in the $5 N_\text{SN} + 11$ dimensional parameter space of $\{ \bm{\phi}_s, E_s, \mu_s \}$ and $\bm{\Theta}$.  We have constructed a Gibbs sampling algorithm that exploits the conditional independence properties of the probabilistic graphical model to generate random moves that efficiently explore the parameter space. It is similar to the \textsc{BayeSN} algorithm of \citet{mandel09,mandel11}, except here the ``data'' are the 3 light curve parameter estimates rather than the full multi-wavelength time series data.  The Gibbs sampler draws from the full set of conditional posterior densities, and does not require tuning of step sizes (which would be required for Metropolis algorithms).   The resulting chains are used to compute numerical summaries of the posterior density.  See Appendix \S \ref{sec:bayesian} and Appendix \S \ref{sec:algorithm} for more details.

In addition to using two separate methods for parameter estimation, maximum likelihood and MCMC sampling, we also implemented these algorithms in independent MatLab and Python codes.  We tested the internal consistency of these codes by simulating data from the model under reasonable hyperparameter values, and then fitting the mock data with each code to verify that the truth was recovered.

\section{Application of Simple-BayeSN to Data}\label{sec:application}

\subsection{The Data Set}

\citet{scolnic15}  cross-calibrated the photometric systems of past SN surveys with the Pan-STARRs system to find one joint solution.  The surveys included are: Pan-STARRS \citep{rest14}, SNLS \citep{betoule14}, SDSS \citep{sako14}, CfA1 \citep{riess99}, CfA2 \citep{jha06}, CfA3 \citep{hicken09a}, CfA4 \citep{hicken12} and CSP \citep{contreras10,stritzinger11}.  The light curve cuts applied to this sample are those used by \citet{betoule14}.  We demonstrate the \textsc{Simple-BayeSN} model by analyzing the nearby sample at low-$z$, where the distances $\mu(z)$ are insensitive to cosmological parameters.  We will present an analysis of the full-$z$ sample in a forthcoming paper.  

In total, there are 248 SN Ia with $0.01 < z < 0.10$ in this sample, with high quality light curves mainly from the CfA and CSP surveys.  The sample includes 29 SN Ia observed both by CfA3/4 and CSP, and we arbitrarily selected the CSP light curves in these cases. The SALT2 light curve model was fit to the optical light curves.  This light curve fitter returns estimates of the optical peak magnitude\footnote{\label{footnote2} \citet{kessler13} describe $m_B$ as ``an effective B-band magnitude is defined to be $m_B = -2.5 \log_{10}(x_0) + 10.635$; this is the observed magnitude through an idealized filter that corresponds to the $B$-band in the rest frame of the SN.''  The overall flux amplitude of the SALT2 fit is $x_0$.  They also show with simulations that the SALT2 $c$ parameter approximates the peak $B-V$ color with  scatter $\approx 0.02$ mag.} $\hat{m}_B$, the peak optical color $\hat{c}$ (corresponding to $B-V$ color), and $\hat{x}_1$ (``stretch''), a measure of the optical light curve shape.  We assign $\bm{d}_s = ( \hat{m}_B, \hat{c}, \hat{x}_1)^T$ as our light curve data.  SALT2 also returns an error covariance matrix for the  light curve fit parameters, which we denote $\bm{W}_s$.

\begin{figure}[t]
\centering
\includegraphics[angle=0,scale=0.43]{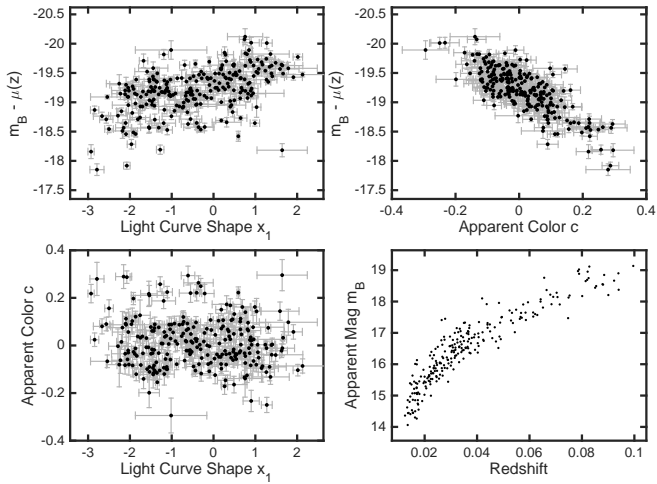}
\caption{\label{fig:fig5} The Data: SALT-II light curve fit parameters and redshifts for 248 SN Ia in the range $0.01 < z < 0.10$.  In the top two panels, we have subtracted the distance modulus expected from the redshift in the fiducial $\Lambda CDM$ cosmology.}
\end{figure}

The sample is cut to a range of light curve shapes of $-3 < \hat{x}_1 < 3$ and a range of colors to $-0.30 < \hat{c} < 0.30$.   The light curve shape cut corresponds approximately to the normal range in optical decline rate $1.6 \gtrsim \Delta m_{15}(B) \gtrsim 0.75$ \citep{hicken09a,kessler09}.  These cuts exclude the very fast declining and red peculiar SN 1991bg-like objects, whose light curves SALT2 was not designed to fit \citep{guy07}.  The median error for the apparent magnitudes $\hat{m}_B$ is $0.066$ mag, the median error for colors $\hat{c}$ is $0.030$ mag, and the median error in the light curve shape $\hat{x}_1$ is 0.121.  Fig. \ref{fig:fig5} summarizes these data.

\subsection{Model Fitting}

\subsubsection{Fitting the Linear Tripp Model}

First, we fit the conventional linear Tripp formula Eq. \ref{eqn:tripp} to the data and estimated the parameters.  We used a Bayesian linear regression that properly accounts for measurement error in the apparent magnitudes $\hat{m}_B$ and covariates ($\hat{c}^\text{app}$, $\hat{x}$), peculiar velocity uncertainties in the absolute magnitudes, as well as the residual variance around the regression $\sigma_\text{res}^2$ (c.f. Appendix \S \ref{sec:fit_tripp}).  We obtained 
\begin{equation}
M_0^{\text{ext},t}        = -19.299 \pm 0.010 \text{ mag},
\end{equation} 
\begin{equation}
\alpha^t     = -0.150 \pm 0.009, 
\end{equation}
\begin{equation}\label{eqn:betaapp}
\beta_\text{app}^t  = 3.01 \pm 0.12, 
\end{equation}
\begin{equation}
\sigma_\text{res}^t = 0.124 \pm 0.009 \text{ mag}.
\end{equation}
The values of $\alpha^t \approx -0.15$ and $\beta^t_\text{app} \approx 3$ are typical for cosmological SN I analyses with the linear Tripp model (e.g. c.f. \citet{rest14}).  The value $\sigma_\text{res}^t = 0.124 \pm 0.009$ mag is also typical, but we caution that this number should \emph{not} be interpreted as the precision of distance estimates in the Hubble diagram.  It corresponds to the portion of the scatter in the Hubble diagram that is unexplained by measurement errors or uncertainties in the SALT2 light curve fits.  Hence, the realized scatter of estimated distance moduli in the Hubble diagram will be larger than this, owing the the light curve fit and peculiar velocity errors.  In \S \ref{sec:distances}, we find that the rms scatter in the Hubble diagram is 0.16 mag for this sample.

\subsubsection{Fitting the Simple-BayeSN model}

Next, we estimated the hyperparameters of the hierarchical model using two numerical methods: optimizing the marginal likelihood and running an MCMC Gibbs sampler.  The results (Table \ref{table:table1}) from the two estimation methods agree quite well, but the posterior estimates typically provide more conservative uncertainty estimates than those obtained from the Fisher matrix evaluated at the maximum likelihood values.

\begin{deluxetable}{lrr}
\tabletypesize{\small}
\tablecaption{Simple-BayeSN Hyperparameter Estimates}
\tablewidth{0pt}
\tablehead{ \colhead{Parameter} & \colhead{Max Likelihood} & \colhead{Posterior}  }
\startdata
$M_0^\text{int}$        & $-19.392 \pm  0.027$    &  $-19.386 \pm 0.029$ \\
$\alpha$    & $-0.154 \pm 0.009$	&  $-0.154 \pm 0.009$ \\
$\beta_\text{int}$  & $+2.252 \pm 0.249$ 	&  $+2.328 \pm 0.262$ \\
$c_0^\text{int}$        & $-0.061 \pm 0.012$	& $-0.059 \pm 0.012$ \\
$\alpha_c^\text{int}$   & $-0.008 \pm 0.005$	& $-0.008 \pm 0.005$ \\
$x_0 $       & $-0.432  \pm 0.074$	& $-0.432 \pm 0.074$ \\
$\sigma_\text{int}$ & $+0.100 \pm 0.013$ &  $+0.104 \pm 0.013$ \\
$\sigma_{c,\text{int}} $  & $+0.065 \pm 0.008$ & $+0.067 \pm 0.009$ \\
$\sigma_x $  & $+1.124 \pm 0.052$ & $+1.134 \pm 0.052$ \\
$R_B $      & $+3.730 \pm 0.308$ &  $+3.758 \pm 0.349$ \\
$\tau$    & $+0.069  \pm 0.012$ &  $+0.068 \pm 0.012$
 \enddata
\tablecomments{\label{table:table1} Hyperparameter Estimates by fitting \textsc{Simple-BayeSN} to the $0.01 < z < 0.10$ SN Ia dataset of SALT-II fit parameters. The Maximum Likelihood Estimates (MLE) are obtained by constrained nonlinear optimization of the log marginal likelihood.  The MLE uncertainties are obtained from the Fisher matrix evaluated at the MLE.  The posterior estimates were obtained via Gibbs sampling and are the posterior mean and standard deviation of the MCMC samples.}
\end{deluxetable}



The slope of the width-luminosity relation $\alpha$ remains the same.  The model estimates the mean and standard deviation of the intrinsic color distribution to be $c_0^\text{int} = -0.06 \pm 0.01$ and $\sigma_{c,\text{int}} = 0.067 \pm 0.009$.  The population average host galaxy dust $E(B-V)$ reddening $\tau$ is non-zero and estimated $\tau = 0.07 \pm 0.01$ mag.   From our posterior estimates, we estimate the population average host galaxy extinction to be $\langle A_V \rangle = \langle \tau (R_B-1) \rangle= 0.184 \pm 0.025$ mag.

\begin{figure}[t]
\centering
\includegraphics[angle=0,scale=0.39]{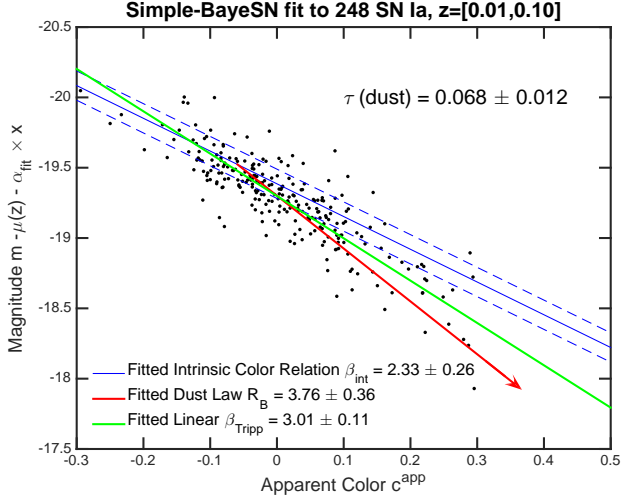}
\caption{\label{fig:fig6} Light curve shape-corrected extinguished absolute magnitudes vs. apparent colors.  The blue line shows the trend of intrinsic absolute magnitude versus intrinsic color with slope $\beta_\text{int}$.  The blue dashed line show the inferred intrinsic scatter around this relation $\sigma_\text{int} = 0.124$ mag.  The red line shows direction of extinction vs. reddening $R_B = 3.8$.  The green line shows the fit from the linear Tripp model.}
\end{figure}

In Figure \ref{fig:fig6}, we show a scatter plot of the light curve shape-corrected extinguished absolute magnitudes versus apparent colors of the sample.  The blue solid line shows the mean trend of intrinsic absolute magnitude versus intrinsic color with slope $\beta_\text{int}$.  The blue dashed lines indicate the inferred intrinsic scatter around this relation $\sigma_\text{int} = 0.10$ mag.  The red vector indicates the direction of dust extinction vs. reddening with the inferred slope $R_B = 3.76$.  The green line shows the trend obtained from the linear Tripp formula fit.

\begin{figure}[t]
\centering
\includegraphics[angle=0,scale=0.41]{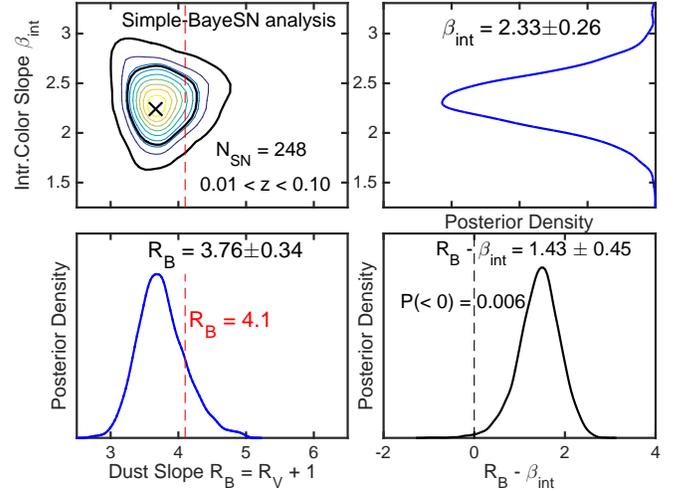}
\caption{\label{fig:fig7} Posterior inferences for the intrinsic color-magnitude slope $\beta_\text{int}$ and the dust law $R_B$ from the MCMC samples.  (top left) The joint density:  The two solid black contours the highest posterior density contours containing the 95\% and 68\% of the posterior probability, and the mode is marked.  (top right) The marginal posterior density of the intrinsic color-magnitude slope $\beta_\text{int}$.  (bottom left)  The marginal posterior density of the dust law slope $R_B$.  It is consistent with the Milky Way Value $R_V = R_B - 1 = 3.1$. (bottom right) The marginal posterior density of the difference between the dust law and the intrinsic color-magnitude slope, $R_B = \beta_\text{int}$.   Their difference is positive and more than $3 \sigma$ from zero.  The tail posterior probability that their difference is less than zero is 0.6\%.} 
\end{figure}

Figure \ref{fig:fig7} shows the joint and marginal posterior probability densities for the hyperparameters $R_B$ and $\beta_\text{int}$.   They indicate that it is highly unlikely $\beta_\text{int}$ or $R_B$ are equal or that either takes on the value of $\beta_\text{app} \approx 3$ found by the linear Tripp model.  In particular the inference of the host galaxy dust law slope $R_B = 3.8 \pm 0.3$ ($R_V = 2.7 \pm 0.3$)  is consistent with the canonical Milky Way average $R_B = 4.1$ $(R_V = 3.1)$.  The intrinsic color-magnitude slope $\beta_\text{int} = 2.33 \pm 0.26$ is significantly ($9 \sigma$) different from zero.
The posterior mean of the difference $R_B - \beta_\text{int}$ is positive and more than $3\sigma$ away from zero, while the tail probability that the difference is less than zero is 0.6\%.

\begin{figure}[t]
\centering
\includegraphics[angle=0,scale=0.42]{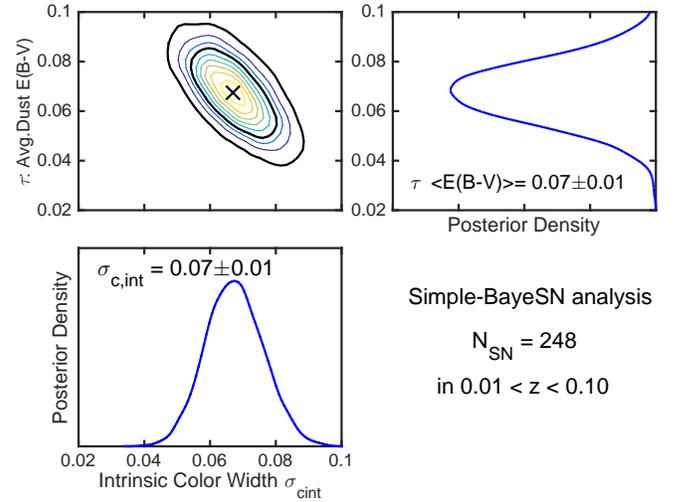}
\caption{\label{fig:fig8} Posterior inferences for population standard deviation width of the intrinsic color $\sigma_\text{c,int}$ and the population average dust reddening $\tau$, computed from the MCMC samples.  (top left)  (top left) The joint density:  The two solid black contours the highest posterior density contours containing the 95\% and 68\% of the posterior probability, and the mode is marked.  (top right) The marginal posterior density of the population mean of the host galaxy dust distribution.  (bottom left)  The marginal posterior density of the intrinsic color width $\sigma_\text{c,int}$. The total apparent color variation comprises approximately equal contributions from intrinsic variation and dust reddening.}
\end{figure}

Figure \ref{fig:fig8} shows the marginal posterior probability densities for the hyperparameters $\sigma_\text{c,int}$ and $\tau$.   Both hyperparameters are well constrained; in particular the population mean of the host galaxy dust distribution is clearly non-zero.  Furthermore, the total apparent color variation comprises approximately equal contributions from intrinsic variation ($\sigma_\text{c,int} \approx 0.07$ mag) and dust reddening ($\tau \approx 0.07$ mag).

\subsection{Latent Distributions}

We explore the implications of our probabilistic model for the latent intrinsic and dust distributions. We examine the probabilistic properties of the distribution of the dusty latent variables that arise from the combination of intrinsic SN Ia variations and host galaxy dust.  These aspects are computed using the fitted values of the hyperparameters $\bm{\hat{\Theta}}$ (Table \ref{table:table1}), and the expressions for the joint distributions $P(M^\text{ext}, c^\text{app}, x, E\, | \, \bm{\hat{\Theta}}_\text{SN})$ and the conditional and marginal distributions derived from it (as described in Appendix \S \ref{sec:app:dustydistr}).

In Figure \ref{fig:fig9}, we compare the data against the joint and marginal model probability distributions of extinguished absolute magnitudes and apparent colors described by the fitted model hyperpameters.  These model distributions are derived from $P(M^\text{ext}, c^\text{app}, x | \, \bm{\Theta})$ as given in Eq. \ref{eqn:psitilde_lkhd}.  The joint distribution of extinguished absolute magnitudes (corrected for light curve shape) and apparent colors accounts for a long redder-dimmer tail attributed to host galaxy dust with a different slope $R_B = 3.8 \pm 0.3$ than the intrinsic color magnitude slope $\beta_\text{int} = 2.3 \pm 0.3$.  The model's marginal apparent color distribution captures the fat, red tail evident in a smoothed kernel density estimate of the apparent color data distribution.  The model also captures an asymmetry in the absolute magnitude distribution.  In these plots, the data distributions are slightly wider than the underlying fitted model distributions due to the effects of estimation errors in the light curve fit parameters and peculiar velocities.

\begin{figure}[t]
\centering
\includegraphics[angle=0,scale=0.39]{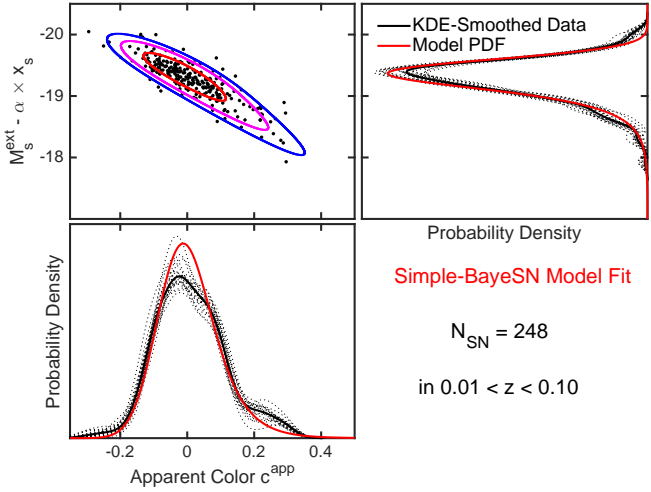}
\caption{\label{fig:fig9} Extinguished absolute magnitudes, corrected for light curve shape, $M^\text{ext}_s - \alpha x_s$ and apparent colors $c^\text{app}$ for the low-redshift $z < 0.10$ sample of 277 SN Ia, compared against the joint and marginal probability densities of the fitted \textsc{Simple-BayeSN} model, derived from $P(M^\text{ext}, c^\text{app}, x | \, \bm{\Theta})$ (Eq. \ref{eqn:psitilde_lkhd}).  (top left) The model joint distribution of $P( M^\text{ext} - \alpha x_s, c^\text{app} | \, \bm{\Theta})$, compared against the data $(m_s - \mu_\text{$\Lambda$CDM}(z_s) - \alpha x_s, c^\text{app}_s)$.  The  blue, magenta, and red isodensity contours enclose approximately $99\%, 95\%$ and $68\%$ of the model probability, respectively.  (top right) Marginal probability of the model (red solid) compared to a smoothed kernel density estimate (KDE) of the extinguished absolute magnitudes (black solid).  The dotted lines are KDEs of 20 bootstrap resampled datasets to reflect the sampling uncertainty in the black solid curve.   (bottom left) The same for the apparent color distribution.  The KDE estimates are slightly wider and shorter than the latent model distributions due to measurement error.}
\end{figure}

\begin{figure}[t]
\centering
\includegraphics[angle=0,scale=0.39]{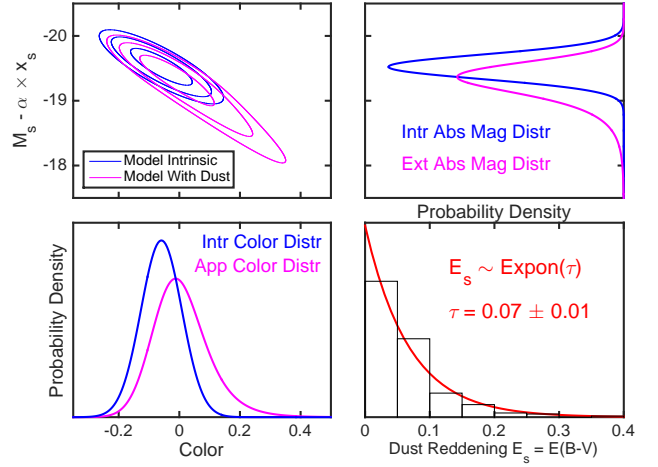}
\caption{\label{fig:fig10} The intrinsic and dusty population distributions of absolute magnitude (corrected for light curve shape) and color, implied by the fitted hyperparameters $\bm{\Theta}$ of our hierarchical model.  (top left)  The model joint distribution of the extinguished absolute magnitudes and apparent colors is shown as the magenta (99\%, 95\%, and 68\% highest density level) contours.  The model's implied joint distribution of intrinsic absolute magnitudes and colors is shown as the blue contours.  (top right) The model's marginal population distribution of intrinsic (blue) and extinguished (red) absolute magnitudes, corrected for light curve shape.  (bottom left) The model's marginal population distribution of intrinsic (blue) and extinguished (red) colors.  (bottom right)  The model's population distribution of host galaxy dust reddening is shown by the red curve, and is assumed to be exponential.  The average $E_s \equiv E(B-V)$ dust reddening is $\tau$.    The histogram shows the distribution of posterior estimates of the dust reddening of individual supernovae $\{E_s\}$ obtained by fitting the dataset.}
\end{figure}

In Figure \ref{fig:fig10}, we illustrate the model intrinsic and dusty population distributions of the absolute magnitude and colors implied by the fitted hyperparameters $\bm{\hat{\Theta}}$.    The model intrinsic distributions are shown in blue.  The population distribution of host galaxy dust reddening $E_s$ is well-described by an exponential distribution with mean $\hat{\tau} = 0.07 \pm 0.01$ (red curve).  The convolution of the intrinsic distribution (blue) with the dust distribution (red) yields the distribution of extinguished magnitudes and apparent colors (magenta).

\begin{figure}[t]
\centering
\includegraphics[angle=0,scale=0.39]{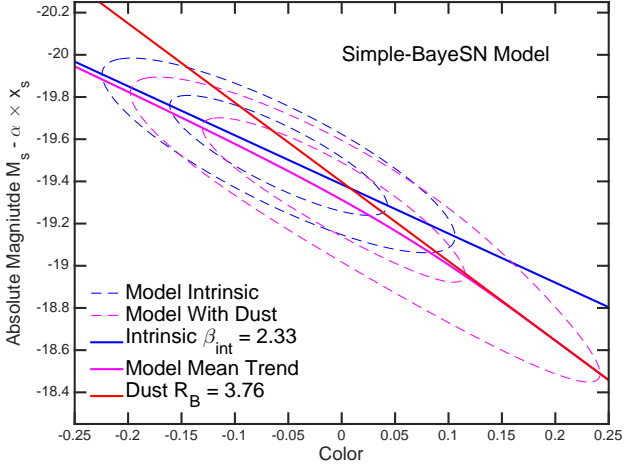}
\caption{\label{fig:fig11} The model joint distributions of (light curve shape-corrected) absolute magnitude and color implied the fitted hyperparameters.   The isodensity contours containing (95\% and 68\%) of the model population distribution are shown as dashed contours.  The blue contours are the model's implied joint distribution intrinsic absolute magnitude and intrinsic color.  The magenta contours depicts the model distribution of extinguished absolute magnitudes and apparent colors.  The mean relation of intrinsic absolute magnitude versus intrinsic color, with slope $\beta_\text{int}$, is the blue solid line, and the red solid line has the slope of the host galaxy dust law, $R_B$.  The nonlinear mean trend of extinguished absolute magnitude (corrected for light curve shape) versus apparent color is shown as the magenta curve.}
\end{figure}

In Figure \ref{fig:fig11}, we further examine the model joint distributions of (light curve shape-corrected) absolute magnitude and color.   The isodensity contours containing approximately 95\% and 68\% of the model population probability are shown as dashed contours.  The blue contours are the model's implied joint distribution intrinsic absolute magnitude and intrinsic color.  The magenta contours depicts the model distribution of extinguished absolute magnitudes and apparent colors.  Also shown are the intrinsic relation between absolute magnitude and intrinsic color, with slope $\beta_\text{int}$ (blue solid line), and a line with the slope of the dust law, $R_B$ (red solid line).  The mean trend of extinguished absolute magnitude (corrected for light curve shape), as a function of apparent color is shown as the magenta curve.  This trend is nonlinear with the following limiting properties.  In the blue (negative) limit of the apparent color distribution, its derivative asymptotes to that of the intrinsic relation $\beta_\text{int}$.  In the red (positive) limit, it acquires the slope of the dust law $R_B$.  The curve smoothly transitions between the two limits at an apparent color value in the red (positive) tail of the intrinsic color distribution.  SN Ia with apparent colors redder (more positive) than this transition are more likely to have been reddened by host galaxy dust rather than to be intrinsically red deviations from the population mean intrinsic color.

\subsection{Distance Estimates}\label{sec:distances}

\begin{figure}[b]
\centering
\includegraphics[angle=0,scale=0.40]{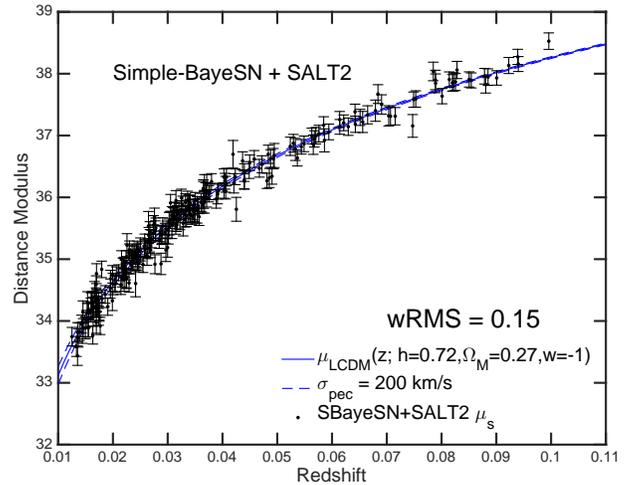}
\caption{\label{fig:fig12} SN Ia distance estimates from \textsc{Simple-BayeSN} applied to the SALT2 light curve fit parameters.  The wRMS for the conventional Tripp formula applied to the same dataset is wRMS $= 0.16$ mag.}
\end{figure}

With the posterior estimates of the hyperparameters, we can calculate photometric distances from the \textsc{Simple-BayeSN} model and compare them to those obtained from the linear Tripp formula.     These are obtained from the probability density $P(\mu_s |\, \bm{d}_s; \bm{\hat{\Theta}})$, for the photometric distance modulus based on the light curve data for each SN $s$, from which we can compute an expected value $\tilde{\mu}_s$ and variance $\tilde{\sigma}_{\mu,s}^2$ (\S \ref{sec:marglkhd}).  This distance uncertainty marginalizes over the tradeoffs between the latent factors of dust reddening-extinction and intrinsic color-magnitude variations.  In Figure \ref{fig:fig12}, we show a Hubble diagram of photometric distance moduli estimates obtained under the fit hyperparameters. 

We compute the precision-weighted root-mean-squared Hubble residuals of the photometric distance moduli relative to the $\Lambda$CDM distance-redshift relation.
\begin{equation}
\text{wRMS}^2 = \left( \sum_{s=1}^{N_\text{SN}} w_s \right)^{-1} \sum_{s=1}^{N_\text{SN}} w_s \left[ \tilde{\mu}_s - \mu_{\Lambda CDM}(z; \bm{\hat{\Omega}} )\right]^2 
\end{equation}
where the precision weights are $w_s^{-1} = \tilde{\sigma}_{\mu,s}^2 + \sigma_{\mu|z,s}^2$.  For the conventional linear Tripp model, scatter of distance modulus estimates about the theoretical distance-redshift relation is about wRMS $= 0.16$ mag.   \textsc{Simple-BayeSN} applied to the same SALT2 parameter estimates yields a minor improvement to wRMS $= 0.15$ mag.  
 
\begin{figure}[t]
\centering
\includegraphics[angle=0,scale=0.38]{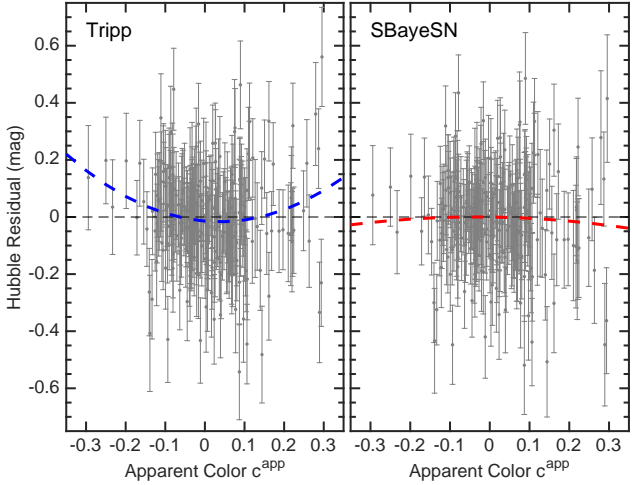}
\caption{\label{fig:fig13} Hubble residuals versus the estimated apparent color $\hat{c}^\text{app}$ for each SN Ia for both the Tripp model (left) and \text{Simple-BayeSN} (right).  We also plot a quadratic fit to the average Hubble residual versus $|\hat{c}^\text{app}|$.   The Tripp model on average overestimates the distance moduli for the SN Ia with very blue and very red apparent colors $|\hat{c}^\text{app}| > 0.2$.   The \textsc{SBayeSN} model reduces the distance biases in the tails of the apparent color distribution by accounting for the specific nonlinear trend caused by the convolution of intrinsic and dust effects.}
\end{figure}

Figure \ref{fig:fig13} shows the Hubble residuals versus the estimated apparent color $\hat{c}^\text{app}$ for each SN for both the Tripp model and \textsc{Simple-BayeSN}.  The conventional linear Tripp model on average overestimates the distance moduli for the SN Ia with very blue and very red apparent colors $|c^\text{app}| > 0.2$.   We fit a simple quadratic curve to the Hubble residuals versus apparent color, $\sum_{p=0}^2 b_p (\hat{c}^\text{app})^p$, and found $b_0 = -0.015 \pm 0.012$, $b_1 = -0.12 \pm 0.11$, and $b_2 = 1.60 \pm 0.65$ (blue dashed curve).  The quadratic coefficient $b_2$ is 2.5$\sigma$ from zero.
\textsc{Simple-BayeSN} reduces the distance bias ($\sim 0.1$ mag) in the tails of the apparent color distribution by fitting for the effective nonlinear trend generated by the convolution of intrinsic and dust effects.  A quadratic fit of the \textsc{Simple-BayeSN} Hubble residuals vs. apparent color yields coefficients consistent with zero.  

\begin{figure}[t]
\centering
\includegraphics[angle=0,scale=0.38]{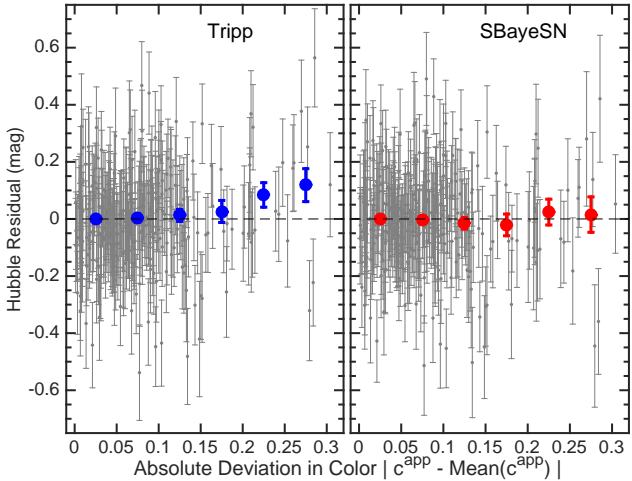}
\caption{\label{fig:fig14} Hubble residuals versus the absolute deviation of apparent color from the mean, $|\Delta \hat{c}^\text{app}| = |\hat{c}^\text{app} - \bar{c}^\text{app}|$  for each SN Ia for both the Tripp model (left) and \text{Simple-BayeSN} (right).  We also plot the  average Hubble residual in each bin of width $0.05$ mag.   The Tripp model systematically overestimates the distance moduli for the SN Ia with very blue and very red apparent colors $|\Delta \hat{c}^\text{app}|> 0.2$.   The \textsc{SBayeSN} model reduces the distance biases in the tails of the apparent color distribution by accounting for the specific nonlinear trend caused by the convolution of intrinsic and dust effects.}
\end{figure}

As shown in \S \ref{sec:motivation}, if the true intrinsic color-magnitude relation is linear and has a slope different from the dust reddening-extinction vector, then the resulting ``dusty'' color-magnitude relation will be a smooth curve.  Attempting to fit this curve with the linear Tripp model results in distance biases that increase as the SN Ia apparent color deviates from the mean, in either direction.  In Figure \ref{fig:fig14}, we plot the Hubble residuals as a function of the absolute deviation of the apparent color from the mean, $|\Delta \hat{c}^\text{app}| = |\hat{c}^\text{app} - \bar{c}^\text{app}|$.  We also compute the mean Hubble residual within bins of width 0.05 mag.  Under the linear Tripp formula, we see that the mean Hubble residual is significantly positive for SN Ia with very red or blue apparent colors $|\Delta \hat{c}^\text{app}|> 0.2$ relative to the mean.  However, the \textsc{Simple-BayeSN} model accounts for the $\beta_\text{int} \neq R_B$ effect, and the Hubble residuals show no trend with $|\Delta \hat{c}^\text{app}|$.

We tested the sensitivity of the model fit to the SN Ia with the most extreme colors, seen on the edges of Fig. \ref{fig:fig13}.  We narrowed our apparent color cut to $|\hat{c}^\text{app}| <0.25$ mag, and refit our model to the remaining 240 SN Ia.  We obtained consistent hyperparameter posterior estimates: $\beta_\text{int}  = 2.4 \pm 0.2$, $R_B = 3.9 \pm 0.4$, and $R_B - \beta_\text{int} = 1.4 \pm 0.5$.  This shows that the fit for the intrinsic slope and dust law is not primarily driven by a few very blue or red SN Ia.  The mean host galaxy reddening hyperparameter decreased to $\tau = 0.05 \pm 0.01$, as expected since the narrower cut removed events with the most dust reddening.

\subsection{Host Galaxy Mass Dependence}\label{sec:hostmass}

In this section, we investigate the dependence of the Hubble residuals on the host galaxy stellar masses, $\mathcal{M}_\text{stellar}$.  \citet{pkelly10} found that the Hubble residuals of SN Ia depended on host mass, after accounting for optical light curve shape and color correlations with luminosity.  This was confirmed in subsequent analyses: the Hubble residuals of SN Ia in more massive galaxies are brighter by about $0.05 - 0.10$ mag \citep{sullivan10, lampeitl10,childress13}.  The SN Ia host galaxies are divided between those with low stellar mass (LM) and high stellar mass (HM).   We adopt the same mass split at $\mathcal{M}_\text{stellar} = 10^{10} \, M_\sun$ used by \citet{betoule14}.

The $0.01 < z < 0.10$ sample contains host mass estimates for 215 SN Ia.  We computed the Hubble residuals obtained via the fitting the linear Tripp model to these SN Ia.  The difference between the mean Hubble residuals of the HM and LM subsamples is $-0.059 \pm 0.028$.  Next, we computed the Hubble residuals obtained with the \textsc{Simple-BayeSN} model fitted to these SN Ia.  The hyperparameter estimates are listed in Table \ref{table:table2}.  The difference between the mean Hubble residuals of the HM and LM subsamples is slightly smaller, $-0.053 \pm 0.028$, with the \textsc{Simple-BayeSN} model.

\begin{deluxetable}{lrlr}
\tabletypesize{\small}
\tablecaption{Fitting Simple-BayeSN with and without \\Host Galaxy Mass Dependence}
\tablewidth{0pt}
\tablehead{ \colhead{Parameter} & \colhead{Estimate} & \colhead{Parameter} & \colhead{Estimate}  }
\startdata
\nodata & \nodata &  $M_{0,\text{LM}}^\text{int}$        & $-19.341 \pm 0.047$   \\
$M_0^\text{int}$  & $19.380 \pm 0.027$ & $M_{0,\text{HM}}^\text{int}$   &  $-19.380 \pm 0.032 $       \\
$\alpha$  & $-0.149 \pm 0.010$ & $\alpha$    & $-0.152   \pm 0.010$  \\
$\beta_\text{int}$  & $+2.322 \pm 0.253$ & $\beta_\text{int}$  & $+2.355  \pm  0.274$ 	\\
$c_0^\text{int}$ & $-0.056 \pm 0.014$ & $c_0^\text{int}$        & $-0.056 \pm 0.013$	 \\
$\alpha_c^\text{int}$ & $-0.006 \pm 0.006$ & $\alpha_c^\text{int}$   & $ -0.007  \pm 0.006$ \\
$x_0 $ & $-0.452 \pm 0.080$ &  $x_0$       & $ -0.452 \pm 0.079$ \\
$\sigma_\text{int}$ & $0.101 \pm 0.013$ &  $\sigma_\text{int}$ & $0.101 \pm  0.014$ \\
$\sigma_{c,\text{int}} $ & $0.071 \pm 0.010$  &  $\sigma_{c,\text{int}} $  & $0.071 \pm 0.009$ \\
$\sigma_x $ & $1.137 \pm 0.058$  &   $\sigma_x $  & $ 1.137 \pm 0.058$  \\
$R_B $ & $+3.740  \pm 0.370$ &  $R_B $      & $ +3.633  \pm 0.417$  \\
\nodata & \nodata & $\tau_{LM} $    & $0.097 \pm 0.027$  \\
$\tau$  & $0.068 \pm 0.014$ &  $\tau_{HM} $  & $0.066 \pm 0.013$ 
 \enddata
 
 
\tablecomments{\label{table:table2} Hyperparameter estimates from fitting the \textsc{Simple-BayeSN} model to the 215 SN Ia in the $0.01 < z < 0.10$ sample with host stellar mass estimates.  The parameters and estimates on the right (left) are obtained by fitting \textsc{Simple-BayeSN} model with (without) host mass step dependence of the intrinsic absolute magnitude offset $M_0^\text{int}$ and the host galaxy dust mean reddening $\tau$. Estimates are the posterior mean and standard deviation of the MCMC samples.}
\end{deluxetable}


Cosmological SN Ia analyses \citep{sullivan11,betoule14} have incorporated this mass dependence into the conventional Tripp formula (Eq. \ref{eqn:tripp}) by modifying the absolute magnitude constant $M_0^\text{ext,t}$ to be a (``mass step'') function of $\mathcal{M}_\text{stellar}$:
\begin{equation}
M_0^\text{ext,t}(\mathcal{M}_\text{stellar}) = \begin{cases} M_{0,\text{LM}}^\text{ext,t}, & \mathcal{M}_\text{stellar} < 10^{10} \, M_\sun \\ M_{0,\text{HM}}^\text{ext,t}, & \mathcal{M}_\text{stellar} \ge 10^{10} \, M_\sun \\ \end{cases}.
\end{equation}
The difference $\delta M_0^\text{ext,t} = M_{0,\text{HM}}^\text{ext,t} - M_{0,\text{LM}}^\text{ext,t}$ is a parameter to be estimated.  This modification accounts for the Hubble residual trend by making the mean \emph{extinguished} absolute magnitude, at zero apparent color $c^\text{app}_s$ and light curve shape $x_s$, depend on host galaxy mass $\mathcal{M}_\text{stellar}$.  

In our model, the hyperparameters  express the intrinsic properties of the SN Ia and the host galaxy dust distribution separately.  To account for a difference in the  \emph{extinguished} absolute magnitude offset between the two host mass classes, one can either make the \emph{intrinsic} absolute magnitude offset $M_0^\text{int}$ a function of host mass, or make the host dust distribution different between the two host mass classes.  In fact, the net effect can be a combination of intrinsic SN Ia and dust population differences between the host mass classes.  The simplest, sensible way to modify \textsc{Simple-BayeSN} to account for an overall offset in \emph{extinguished} absolute magnitude is to allow both $M_0^\text{int}$ and $\tau$ to depend on host mass.
\begin{equation}
M_0^\text{int}(\mathcal{M}_\text{stellar}) = \begin{cases} M_{0,\text{LM}}^\text{int}, & \mathcal{M}_\text{stellar} < 10^{10} \, M_\sun \\ M_{0,\text{HM}}^\text{int}, & \mathcal{M}_\text{stellar} \ge 10^{10} \, M_\sun \\ \end{cases}
\end{equation}
\begin{equation}
\tau(\mathcal{M}_\text{stellar}) = \begin{cases} \tau_{LM}, & \mathcal{M}_\text{stellar} < 10^{10} \, M_\sun \\ \tau_{HM}, & \mathcal{M}_\text{stellar} \ge 10^{10} \, M_\sun \\ \end{cases}
\end{equation}

We have modified our Gibbs sampler to sample the posterior distribution in the expanded parameter set and ran it on our sample to train the model and estimate the hyperparameters.   The posterior mean estimates of the hyperparameters are shown in Table \ref{table:table2}.  We recomputed distances and the Hubble residuals obtained from the trained \textsc{Simple-BayeSN} model with host mass dependence.   After including intrinsic $M_0^\text{int}$ and dust $\tau$ host mass dependences in the model to compute distances, the difference between the mean Hubble residuals of the HM and LM subsamples is $-0.005 \pm 0.028$ mag.

\begin{figure}[t]
\centering
\includegraphics[angle=0,scale=0.40]{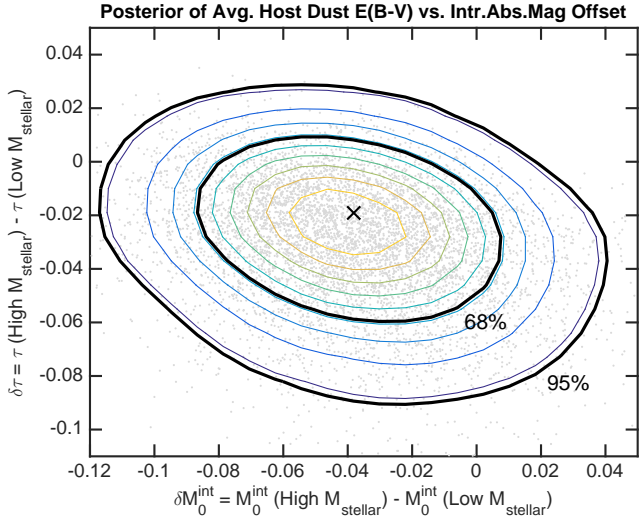}
\caption{\label{fig:fig15} Joint posterior density of the relative intrinsic absolute magnitude offset $\delta M_0^\text{int} = M_{0,\text{HM}}^\text{int}  -M_{0,\text{LM}}^\text{int}$ and the relative difference in the population average host galaxy dust reddening $\delta \tau \equiv \tau_{HM} - \tau_{LM}$, between SN Ia subsets in high or low stellar mass host galaxies.}
\end{figure}

In Figure \ref{fig:fig15}, we show the joint posterior density of the difference in the average host galaxy dust reddening $\delta \tau \equiv \tau_{HM} - \tau_{LM}$ and the intrinsic absolute magnitude offset $\delta M_0^\text{int} = M_{0,\text{HM}}^\text{int}  -M_{0,\text{LM}}^\text{int}$.  The peak of this distribution favors a solution that  explains the Hubble-residual mass step partially by intrinsic SN Ia $\delta M_0^\text{int} = -0.04$ mag, and partially by dust distribution $\delta \tau \approx -0.02$ mag differences.  However, as the size of this low-$z$ sample is small, the uncertainties are wide; $(\delta \tau = 0, \delta M_0^\text{int} = 0)$ falls just inside the 95\% highest posterior density contour, and the effects may be sensitive to the choice of the mass split.  Future application of this model to a large high-$z$ cosmological sample will yield more robust constraints on these host galaxy dependencies.

\section{Discussion}\label{sec:discussion}

Our \textsc{Simple-BayeSN} model describes the observed ``dusty'' distribution of the extinguished absolute magnitudes and apparent colors of SN Ia as arising from the combination of intrinsic SN Ia color-magnitude variations (independent from light curve shape) and the distribution of host galaxy dust reddening-extinction.  The probabilistic convolution of these two distributions generically results in a non-linear dusty apparent color-magnitude trend (Figs. \ref{fig:fig4}, \ref{fig:fig11}).  Previous empirical studies have noted evidence for trends of Hubble residuals with respect to apparent color in the conventional linear Tripp analysis, indicating that SN Ia with bluer or redder apparent colors may follow different effective color-magnitude slopes.  \citet{sullivan11} analyzed SNLS3 SN Ia data using the linear Tripp formula and found $\beta^t_\text{app} = 3.097\pm 0.094$, but noted a possible trend of Hubble residuals blueward of $c^\text{app} < 0.15$, suggesting bluer SN preferred a shallower effective color-magnitude slope.  Puzzlingly, when they separately analyzed subsamples split at $c^\text{app} = 0$, they found each $\beta^t_\text{app} \approx 3.8$ to be higher than the global slope.  

\citet{ganeshalingam13} analyzed the combination of the Lick Observatory Supernova Search (LOSS) SN Ia sample \citep{ganeshalingam10} with other low-$z$ and high-$z$ samples, using SALT2 and the linear Tripp method.  They estimated a typical $\beta_\text{app}^t \approx 3.17 \pm 0.08$, but found significant positive Hubble residuals at apparent colors $c^\text{app} < -0.10$, indicating that a shallower apparent color-magnitude trend (smaller $\beta_\text{app}$) would be a better fit to the bluer SN Ia.  They also noted a slight positive deviation in the average Hubble residuals for $c^\text{app} > 0.3$, but it was not statistically significant.  \citet{scolnic14} analyzed a compilation of SNLS3, SDSS-II and nearby samples with the linear Tripp relation and found that the Hubble residuals for SN Ia with $c^\text{app}$ bluer (more negative) and redder (more positive) than zero favored two different slopes, $\beta_\text{app} = 1.68 \pm 0.38$ and $\beta_\text{app} = 3.22 \pm 0.29$, respectively.

Analyzing the Union 2.1 data compilation, \citet{suzuki12} find a shallower $\beta_\text{app} = 1.3 \pm 0.3$ for bluer SN ($c^\text{app} < 0.05$), and a steeper slope $\beta_\text{app} = 2.77 \pm 0.09$ for redder SN ($c^\text{app} > 0.05$).  \citet{rubin15} included a broken-linear apparent color-magnitude relation in their Bayesian UNITY model and found $\beta_\text{app} = 0.6 \pm 0.3$ for blue $c^\text{app} < 0$, and $\beta_\text{app} = 2.8 \pm 0.2$ for red $c^\text{app} > 0$, when they applied it to analyze Union 2.1.  Notably, the two different apparent blue and red slopes are both much less than the $R_B = 4.1$ of normal MW interstellar dust.

Fitting for different apparent color-magnitude slopes $\beta_\text{app}$ for apparently blue and red SN Ia is a useful empirical test for non-linearity of effective color-magnitude trend.  However, as we pointed out in \S \ref{sec:comp_tripp}, the incorporation of \emph{any} (linear or nonlinear) functional form $m_s - \mu_s = M_s^\text{ext} = f(c_s^\text{app}, x_s)$, directly between extinguished absolute magnitudes and apparent light curve parameters, within a statistical model for SN Ia data,  would inherit the same conceptual flaw as the conventional linear Tripp formula.  It would not properly weigh the probabilities of the different random combinations of intrinsic ($M^\text{int}, c^\text{int}$) and dust ($A_B, E$) that could have produced a given ($M^\text{ext}, c^\text{app}$), and it would fail to capture the essential probabilistic nature of the latent astrophysical processes underlying the data.   

By probabilistically deconvolving the SN Ia data into intrinsic variations and host galaxy dust components, we estimate a dust law with $R_B = 3.8 \pm 0.3$, consistent with the normal MW value of $R_B = 4.1$.  This is consistent with recent SN Ia analyses that do not use the Tripp formula.  \citet{burns14}, analyzing the optical-NIR colors of nearby SN Ia, found that those with low reddening $E(B-V) < 0.3$ are consistent with dust reddening with $R_B = 4.1$ \citep[c.f.][]{phillips12}.  \citet{mandel11}, accounting for intrinsic SN Ia covariance over phase and wavelength with the \textsc{BayeSN} hierarchical optical-NIR light curve model, found an $R_B \approx 3.8$ in the limit of low reddening.  Correlating spectral equivalent widths with intrinsic photometric variability, \cite{chotard11} estimate $R_B = 3.8 \pm 0.3$.  \citet{sasdelli16} developed a new technique to use principal components of SN Ia spectra to predict the intrinsic $B$ light curves and $B-V$ color curves of nearby SN Ia.  By comparing these predictions against the observed light curves, they were able to deduce dust reddening and extinction and estimated $R_B = 3.8 \pm 0.3$.   These studies have leveraged additional observational data (NIR photometry or spectra) to reach their conclusions.  In this work, we have obtained a consistent and sensible estimate of $R_B$ using only optical photometric data fit with the same SALT2 light curve model used for cosmological SN Ia studies.

Our results for the dust law $R_B = 3.8 \pm 0.3$ are in quantitative agreement with the results of \citet{scolnic14}, who found a $\beta \approx 3.7$ when attributing the residual scatter to solely to color $\sigma_{c_r}$ within the \textsc{SALT2mu} framework \citep{marriner11}.  One may be tempted to identify their residual scatter in color $c_r$ (Eq. \ref{eqn:salt2mu}) to \textsc{Simple-BayeSN}'s intrinsic color $c^\text{int}$.  However, our approaches are conceptually different.   \citet{marriner11} introduced the residual scatter matrix for additional dispersion in the light curve parameters that, in effect, acts in the same way as measurement errors around the primary magnitude, light curve shape and color relations of the usual Tripp model.  In \textsc{Simple-BayeSN}, the intrinsic SN Ia distribution is fundamental: supernovae have physical intrinsic colors and luminosities \emph{before} any dust reddening-extinction or measurement occurs (\S \ref{sec:motivation}).

There are also important practical differences.  Before applying \textsc{SALT2mu} to fit for the Tripp coefficients, it is necessary to first assume the proportions of residual scatter attributed to luminosity and color, since it does not itself estimate them from the data.   Further, \citet{scolnic14} assumed the residual scatter in color is uncorrelated with luminosity.  \textsc{SALT2mu} minimizes a ``$\chi^2$'', a simplistic method with known frequentist biases in linear regression when dealing with scatter in the covariates (c.f. \S \ref{sec:fit_tripp}).  

In contrast, \textsc{Simple-BayeSN} is a hierarchical Bayesian model that estimates the hyperparameters of the intrinsic SN Ia and dust component populations from the observed data.  The host galaxy dust distribution is given a physically motivated form that only allows positive extinction.  We find $\beta_\text{int} = 2.3 \pm 0.3$, indicating that the intrinsic colors of SN Ia are correlated with their intrinsic luminosity, for a given light curve shape.  Hence, the ``model'' $c_\text{mod}$ and ``residual'' $c_r$ color components in Eq. \ref{eqn:salt2mu} in the \textsc{SALT2mu} analysis of \citet{scolnic14}  do not exactly correspond to the dust reddening $E(B-V)$ and intrinsic color $c^\text{int}$, respectively, in the \textsc{Simple-BayeSN} framework presented here.  However, is it likely that combinations of the former map into  combinations of the latter.

Our statistical analysis indicates significant intrinsic color-luminosity variation ($\beta_\text{int} = 2.3 \pm 0.3$, $\sigma_{c,\text{int}} = 0.07 \pm 0.01$) independent from light curve shape.  This variation may result from the effects of observational viewing angle into asymmetric SN Ia explosions \citep{krw09, maeda11}.  Theoretical simulations of multi-dimensional asymmetric delayed-detonation SN Ia explosions by \citet*{krw09} follow an intrinsic color-luminosity slope of $\beta_\text{int} = 4.45$, controlling for light curve shape (in the absence of dust).  Reaching quantitative agreement with the shallower slope found by our statistical model will be a challenge for theoretical models of SN Ia explosion physics.

We have used the SALT2 SN Ia spectral template model to fit the optical photometric time series to obtain estimates of the peak apparent $m_B$, $c$ ($B-V$ color), and light curve shape $x_1$ for each SN Ia.   Using \textsc{Simple-BayeSN} to analyze these derived parameter estimates and find two latent trends, $\beta_\text{int}$ and $R_B$.  However, internal to SALT2, there is a single, empirically-derived average color law $CL(\lambda)$, describing the color variation in the SN Ia spectral energy distribution independent from light curve shape.  Our results suggest that an improved SN Ia SED model should internally account for two physical sources of chromatic variation: dust reddening and intrinsic color variations.

Previous studies have found a correlation between the peak intrinsic $B-V$ color and the optical decline rate, with a slope of $\approx +0.10$ against $\Delta m{15}(B)$ \citep{phillips99,altavilla04,folatelli10}.  This roughly translates to a slope of $\approx -0.014$ against SALT2 stretch $x_1$.  Using SALT2 fit parameters, our estimate of the intrinsic color-light curve shape slope is $\alpha_c^\text{int} = -0.008 \pm 0.005$ (Eq. \ref{eqn:intrinsic2}, Table \ref{table:table1}) is not inconsistent with those estimates.  However, it may be a shallower slope because internally the SALT2 model only has one color law and the model's peak $B-V$ color has a negligible dependence on light curve shape $x_1$ \citep{kessler13}.  Applying \textsc{Simple-BayeSN} to the light curve estimates from an improved fitting method may tighten the constraints on this and other key supernova parameters.

Our estimated $R_B = 3.8 \pm 0.3$ may in fact be somewhat biased low due to observational selection effects, which we have not modeled here.  For any given apparent color $c^\text{app}$, supernovae dimmer than the flux limits of nearby supernova searches will not be found and followed up.  Hence, SN Ia with extinguished absolute magnitudes near the bottom portion of Fig. \ref{fig:fig6} would not be found, and their absence may cause the estimated $R_B$ vector to be slightly shallower.  Although this mainly affects dust-reddened SN Ia, this bias is not the same as a cut in apparent color.  Proper modeling the selection effects of nearby surveys is challenging, but such a correction would to tend to push $R_B$ closer to the MW mean $4.1$.

Another potentially important interplay between our model and observational biases would occur at high-$z$, where selection effects are strong.  The combination of a nonlinear apparent color-magnitude curve with high-$z$ selection biases may contribute to the apparent decrease in the estimated $\beta_\text{app}$ with redshift (``$\beta$ evolution''). \citet{kessler09}'s analysis of the SDSS-II data found that when the SALT2 parameters were fit with the Tripp formula to  subsamples binned by redshift, the resulting estimates of $\beta_\text{app}$ tended to decrease significantly at $z > 0.6$.   Using an updated SALT2, \citet{guy10} also found a decreasing estimated $\beta_\text{app}$ at higher $z$, but did not conclude that it was real, due to the possibility of systematic errors in the estimation of SN Ia color and its uncertainties confounding the determination of $\beta_\text{app}$ at high-$z$.  While \citet{betoule14} did not report any $\beta$ evolution, recent analyses of their JLA compilation by \citet{li16} and \citet{shariff16} have claimed decreasing $\beta_\text{app}$ at high-$z$.  For example, \citet{shariff16} report an unexplained drop from $\beta_\text{app} = 3.1 \pm 0.1$ by $\Delta \beta_\text{app} = -1.1 \pm 0.2$ at $z \approx 0.66$.  

We demonstrated in \S \ref{sec:motivation} that, if the data is generated from intrinsic variation and dust distributions with different $\beta_\text{int}$ and $R_B$, then their probabilistic convolution results in a nonlinear mean apparent color-magnitude trend.   When the observed data is fit with the linear Tripp formula, it yields an estimated $\beta^t_\text{app}$ that approximates the derivative of this curve near the mean of the apparent color distribution (Fig. \ref{fig:fig4}).  At high-redshifts, the dimmer, and thus redder and dustier SN Ia, are less likely to be found and included in the sample.  This will cause the mean apparent color of the observed sample to shift to bluer (more negative) values from low to high redshift.  The observed high-$z$ sample will be, in effect, less dusty causing the apparent color-magnitude curve to be shallower overall.  The blue shift in the mean apparent color at high redshift will cause the $\beta^t_\text{app}$ to approximate the derivative of the underlying curve at a bluer (more negative) mean apparent color, where the derivative is closer to the intrinsic slope $\beta_\text{int}$ (Fig. \ref{fig:fig11}). The net result will naturally cause the Tripp estimator of $\beta^t_\text{app}$ to obtain a lower value at high-$z$ relative to low-$z$.   

In the context of our results,  our estimate $\beta^t_\text{app} = 3.01 \pm 0.12$ (Eq. \ref{eqn:betaapp}) is consistent with the $\beta^t_\text{app} \approx 3$ found at $z \lesssim 0.7$ by \citet{li16} and \citet{shariff16}, and our $\beta_\text{int} = 2.3 \pm 0.3$ is consistent with $\beta^t_\text{app} \approx 2$ they find at higher redshifts.  This is consistent with our understanding above that the SN Ia at high-$z$ will be on average bluer and less dusty due to selection bias, and thus should adhere closer to the shallow intrinsic color-luminosity trend $\beta_\text{int}$.  This explanation rests on the relative proportions of intrinsic color to dust reddening, in the observed sample, changing with redshift due to selection bias, and does \emph{not} require either the intrinsic slope $\beta_\text{int}$ or dust law $R_B$ themselves to ``evolve'' with redshift.  Disentangling this generic prediction of our model from the possible systematic errors particular to the fitting of noisy high-$z$ SN Ia light curve data will be a challenge.

Several empirical studies have looked for astrophysical systematic effects by examining the distribution of SN Ia parameters, or the Hubble residuals, as a function of host galaxy properties or local environment of the SN \citep[e.g.][]{pkelly10,lampeitl10,sullivan10,kelly15}.  For example, \citet{sullivan11} find a $4.4\sigma$ difference in $\beta^t_\text{app}$ between SN Ia host galaxies with low and high stellar mass.  They note that the estimated $\beta^t_\text{app}$ is likely conflating dust effects with intrinsic variations, and hypothesize that this difference is tied to the relative dustiness of host galaxies as a function of stellar mass and star formation rate.  Because of the confounding of intrinsic variation and dust, the Tripp formula is a blunt tool to examine these hypotheses.

In contrast, our framework is well-suited to these kinds of analyses because we have separately modeled the latent components of the data attributed to intrinsic variation and host galaxy dust, and parameterized their properties (e.g. $\beta_\text{int}$, $R_B$, $\sigma_\text{c,int}$, $\tau$).  This provides a richer vocabulary with which to investigate the astrophysical connections between the SN observables and the properties of the progenitor, local environments or host galaxies.  For example, in \S \ref{sec:hostmass} we incorporated the host galaxy ``mass step'' into our model in a new way: we allow a combination of the intrinsic SN Ia brightness and host galaxy dustiness to vary with host stellar mass.  This approach may lead to insights into the physical origin of the observed effect, e.g. whether it is caused by the the host galaxy dust properties and/or the physics of SN Ia progenitors.  This is just one simple example of how our framework can be used to investigate astrophysical systematics.  Future studies could investigate the dependence of our intrinsic and dust parameters on star formation rate, host galaxy stellar mass, metallicity and morphology.

\section{Conclusion}\label{sec:conclusion}

The Tripp formula (Eq. \ref{eqn:tripp1}) is widely used in conventional analyses of cosmological SN Ia light curve data.  However, it is also simplistic: by directly regressing \textit{extinguished} absolute magnitudes against \textit{apparent} color, it fails to take into account that both factors comprise the physically distinct effects of intrinsic SN Ia variation and extrinsic host galaxy dust.  This shortcoming has led to estimates of the apparent color-magnitude slope $\beta_\text{app}$ that are puzzlingly smaller than the normal MW interstellar dust reddening-extinction law $R_B = 4.1$.

To address this, we have constructed a hierarchical Bayesian model (\textsc{Simple-BayeSN}) describing the ``dusty'' distribution of extinguished absolute magnitudes and apparent colors as arising from two population distributions: an intrinsic SN Ia distribution, and a host galaxy dust distribution.  The intrinsic distribution includes the dependence of intrinsic luminosity on light curve shape and intrinsic color (Eq. \ref{eqn:intrinsic1}).  It allows for a non-zero trend of intrinsic absolute magnitude vs. intrinsic color with slope $\beta_\text{int}$, controlling for light curve shape.  It also models intrinsic color variations that can be uncorrelated  with light curve shape (Eq. \ref{eqn:intrinsic2}).  The host galaxy dust distribution includes a law parameter, $R_B$, characterizing the direction of the dust extinction-reddening vector.   This model provides a more physical decomposition of the sources of variation underlying the SN Ia data.  Inference with this model in effect performs a probabilistic deconvolution of the observed SN Ia measurements to estimate the characteristics of the underlying intrinsic and dust component distributions.

Analyzing the optical light curve fit data from compilation of 248 nearby SN Ia ($z  < 0.1$), we find that fitting the Tripp formula obtains $\beta^t_\text{app} = 3.0 \pm 0.1$, significantly less than the $R_B = 4.1$ of normal MW dust, but consistent with the findings of recent cosmological analysis.
In contrast, \textsc{Simple-BayeSN}, by modeling the data as probabilistic convolution of intrinsic and dust components, finds a non-zero intrinsic color-magnitude slope $\beta_\text{int} = 2.3 \pm 0.3$ and a dust law slope of $R_B = 3.8 \pm 0.3$.  The slope of the dust law is consistent with the average value for normal Milky Way interstellar dust $R_B = 4.1$.  The width of the intrinsic $B-V$ color distribution is found to be $\sigma_{c,\text{int}} \approx 0.07$ mag, while the average dust $E(B-V)$ reddening of the sample is $\tau \approx 0.07$ mag.

Since $\beta_\text{int} \neq R_B$ (at $3\sigma$), the convolution of the intrinsic SN Ia color-magnitude distribution with the dust reddening-extinction distribution results in a non-linear curve of extinguished absolute magnitude vs. apparent color (Fig. \ref{fig:fig4} \& \ref{fig:fig11}).  The conventional linear Tripp formula approximates this curve near the bulk of the empirical apparent color distribution of the samples.  It obtains a linear slope that approximates the slope of a tangent to this curve near the mean apparent color.  This results in $\sim 0.1$ mag overestimates of photometric distance moduli in the tails of the apparent color distribution.  This systematic bias vanishes when we use our model to account for the distinct effects of intrinsic variation and host galaxy dust.

 As photometric calibration uncertainties become better understood and controlled \citep{scolnic15}, astrophysical systematics caused by incorrect modeling of SN Ia color-luminosity effects will become a major limiting factor for dark energy constraints from SN Ia.  Future research will evaluate the relative impacts of these systematics on the constraints on $w$ from the cosmological SN Ia sample.  Since the Tripp formula is also used for calibrating nearby SN Ia on the absolute distance scale, similar color-dependent systematic errors may impact the inference of $H_0$.  However, the effect on the $H_0$ estimate is likely to be small, as it is mainly influenced by the measurement of the average luminosity of the low-$z$ SN Ia, whereas these biases are most pronounced in the color tails.

There are many potential directions for applying and extending the \textsc{Simple-BayeSN} framework.  We will apply this model for the cosmological analysis of SN Ia over the full range of redshifts, accounting for systematics and selection effects, for determination of the cosmological parameters.   While we have applied \textsc{Simple-BayeSN} to model intrinsic and dust effects using parameters derived from optical photometric light curve fits, we can incorporate other useful information that is likely to improve inferences.  For example, additional measurements from NIR light curves \citep[e.g.][]{friedman15} is likely to improve constraints on dust and distances.  We can also extend the model to incorporate spectroscopic information.  For example, the spectroscopic expansion velocity-color relation \citep[VCR,][]{foleykasen11} can be modeled by adding velocity as an additional variable that may correlate with intrinsic color and luminosity.  Non-Gaussian distributions of the intrinsic parameters can be tested \citep{mandel14}.    The \textsc{Simple-BayeSN} hyperparameters could be used to explore astrophysical systematics related to the dependence of intrinsic SN Ia properties (e.g. $\beta_\text{int}$ and $\sigma_\text{c,int}$) and dust ($\tau, R_B$) with properties of the host galaxy and local environment.   We will incorporate this model into SNANA \citep{snana09} to use in realistic simulations of SN Ia surveys.

Precise and accurate SN Ia distance estimates are essential to the success of current or future cosmological surveys such as the Dark Energy Survey (DES), the Large Synoptic Survey Telescope (LSST) survey, and WFIRST.   \textsc{Simple-BayeSN} is both a conceptual advance and practical improvement in the proper statistical modeling, inference and understanding of the intrinsic SN Ia variations and host galaxy dust effects underlying these measurements.

\acknowledgements

We thank Saurabh Jha, Rick Kessler, Pat Kelly, Xiao-Li Meng, David Spergel, and Roberto Trotta for useful discussions.  Supernova cosmology at the Harvard College Observatory is supported in part by National Science Foundation grants AST-156854,  AST-1211196, and NASA grant NNX15AJ55G.  This manuscript is based upon work supported by the National Aeronautics and Space Administration under Contract No. NNG16PJ34C issued through the WFIRST Science Investigation Teams Program.  R.J.F. and D.S. were supported in part by NASA grant 14-WPS14-0048. R.J.F.'s UCSC group is supported in part by NSF grant AST-1518052 and the Alfred P. Sloan Foundation.  D.S. acknowledges support from KICP and from NASA through Hubble Fellowship grant HST-HF2-51383.001 awarded by the Space Telescope Science Institute, which is operated by the Association of Universities for Research in Astronomy, Inc., for NASA, under contract NAS 5-26555.  H.S. was supported by a Marie-Skodowska-Curie RISE (H2020-MSCA-RISE-2015-691164) Grant provided by the European Commission.  This work was supported by Grant ST/N000838/1 from the Science and Technology Facilities Council (UK).

\appendix

\section{The Dusty SN Ia Distribution}\label{sec:app:dustydistr}

The joint probability distribution of the dusty SN Ia parameters $P(M^\text{ext}, c^\text{app}, x |\, \bm{\Theta})$ can be derived from the model assumptions for the intrinsic SN Ia population distribution (\S \ref{sec:intrinsic_model}) and the dust population model (\S \ref{sec:dust_model}).  This depends on the their respective hyperparameters: $\bm{\Theta}_\text{SN}$, and $\bm{\Theta}_\text{dust} = (R_B, \tau)$.  Let $\bm{\tilde{\psi}} = (M^\text{ext}, c^\text{app}, x_s)^T$ be a vector of the dusty parameters of a SN Ia, modified by extinction and reddening.  This is related to the intrinsic SN Ia parameters $\bm{\psi}$ via $\bm{\tilde{\psi}} = \bm{\psi} + \bm{e}_E E$, where $\bm{e}_E = \bm{e}_2 + R_B \bm{e}_1$.   The joint distribution of $\bm{\tilde{\psi}}$ and dust reddening $E$ is
\begin{equation}\label{eqn:psitildeE_lkhd}
\begin{split}
P(\bm{\tilde{\psi}}, E |\, \bm{\Theta}) &= P(\bm{\psi} = \bm{\tilde{\psi}} - \bm{e}_E E | \, \bm{\Theta}_\text{SN}, R_B) P( E | \,\tau).
\end{split}
\end{equation}
The marginal probability distribution of $\bm{\tilde{\psi}}$ is a convolution of the intrinsic SN Ia population distribution and the dust distribution for a given set of model hyperparameters $\bm{\Theta} = \{ \bm{\Theta}_\text{SN}, R_B, \tau \}$, obtained by this integral:
\begin{equation}\label{eqn:psitilde_lkhd}
\begin{split}
P( \bm{\tilde{\psi}} | \, \bm{\Theta}) &=  \int  dE \, P(\bm{\tilde{\psi}}, E | \bm{\Theta})
\end{split}
\end{equation}
From this, one can compute the conditional probability $P(M^\text{ext} | \, c^\text{app}, x, \bm{\Theta})$ and then the mean trend of extinguished absolute magnitude versus apparent color and light curve shape, $\mathbb{E}(M^\text{ext} | \, c^\text{app}, x, \bm{\Theta})$.  The joint distribution of the light curve shape-corrected extinguished absolute magnitude $M^\text{ext} -\alpha x$ and apparent color is mathematically equivalent to evaluating the above expressions with $x = 0$.

\section{Bayesian Inference}\label{sec:bayesian}

\subsection{Global Posterior Probability Density}

To estimate all the parameters and hyperparameters, we compute the joint posterior density.  For one SN $s$, the joint probability of the data $\bm{d}_s $ and latent parameters $\bm{\phi}_s, E_s$, and $\mu_s$ given redshift $z_s$, the hyperparameters $\bm{\Theta} = (\bm{\Theta}_\text{SN}, \bm{\Theta}_\text{dust})$ and cosmological parameters $\bm{\Omega}$ is:
\begin{equation}\label{eqn:full_lkhd}
\begin{split}
P( \bm{d}_s, \bm{\phi}_s, E_s, \mu_s | \, z_s; \bm{\Theta}, \bm{\Omega} )&= P( \bm{d}_s | \bm{\phi}_s ) \times P(E_s | \tau)  \times P(\mu_s | \, z_s; \bm{\Omega})  \\
 &\times P( \bm{\psi}_s = \bm{\phi}_s - \bm{e}_1 \mu_s - \bm{e}_E E_s |\, \bm{\Theta}_\text{SN}, R_B ) 
\end{split}
\end{equation}

The global posterior probability density of all parameters and hyperparameters, conditioning on all the light curve data $\mathcal{D} = \{ \bm{d}_s \}$ and the redshifts $\mathcal{Z} = \{ z_s \}$ of the sample with $\bm{\Omega}$ fixed is:
\begin{equation}\label{eqn:globalposterior}
P( \{ \bm{\phi}_s, E_s, \mu_s \}; \bm{\Theta} | \, \mathcal{D}, \mathcal{Z}; \bm{\Omega} ) \propto \Bigg[ \prod_{s=1}^N P( \bm{d}_s, \bm{\phi}_s, E_s, \mu_s | \,z_s; \bm{\Theta}, \bm{\Omega} ) \Bigg] P(\bm{\Theta})
\end{equation}
assuming the conditional independence of the individual SN Ia. The joint posterior of all latent variables $\{ \bm{\phi}_s, E_s, \mu_s \}$ and hyperparameters $\bm{\Theta}$ can then be computed by drawing samples from this distribution (\S \ref{sec:algorithm}).  

Joint inference of all parameters including the cosmological parameters $\bm{\Omega}$ could be computed from
\begin{equation}
P( \{ \bm{\phi}_s, E_s, \mu_s \}; \bm{\Theta, \bm{\Omega} } | \, \mathcal{D}, \mathcal{Z} ) \propto P( \{ \bm{\phi}_s, E_s, \mu_s \}; \bm{\Theta} | \, \mathcal{D}, \mathcal{Z}; \bm{\Omega} ) P(\bm{\Omega})
\end{equation}
where $P(\bm{\Omega})$ includes constraints from other data (e.g. CMB or BAO).  However, we do not do this in this paper, since we are restricting our analysis to the low-$z$ sample with $\bm{\Omega}$ fixed.

\subsection{Probabilistic Graphical Model}

\begin{figure}[h]
\centering
\includegraphics[angle=0,scale=0.6]{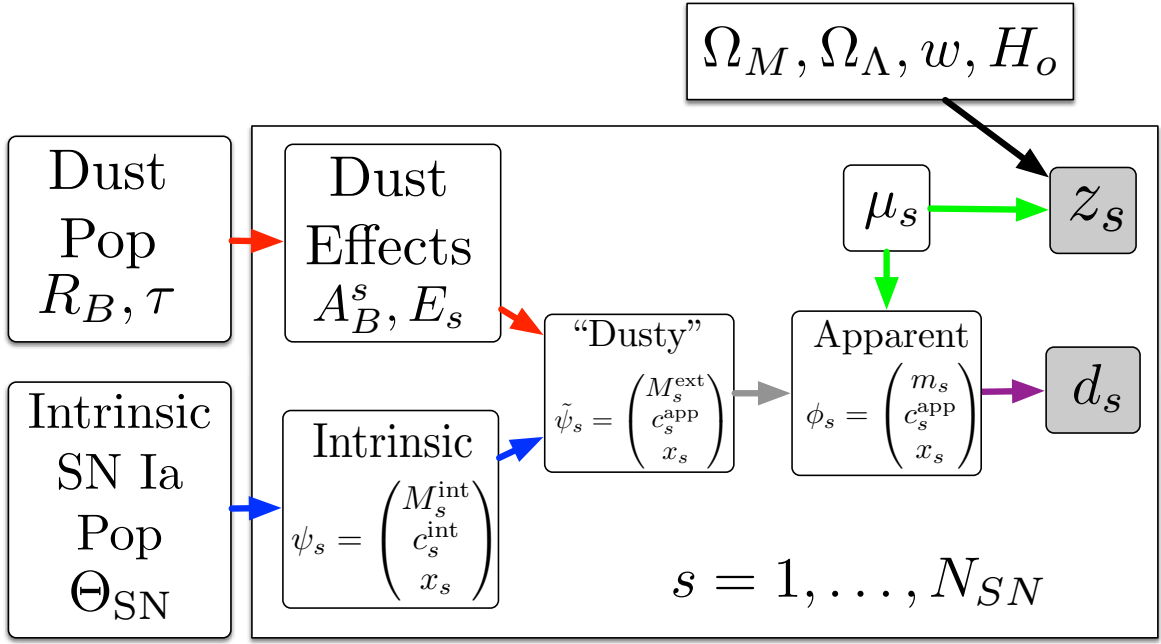}
\caption{\label{fig:fig16} A probabilistic graphical model describing \textsc{Simple-BayeSN}.  The open boxes represent unknown parameters and hyperparameters, the shaded boxes represent the observed data, and the arrows indicate relations of conditional probability between the parameters, hyperparameters and data.  The arrow colors heuristically represent the multiple sources of randomness and uncertainty underlying the data: intrinsic variation (blue), dust (red), distances and peculiar velocities (green), measurement errors (purple) and cosmological parameters (black). One can read the graph as a describing the generative process for the data.}
\end{figure}

In Figure \ref{fig:fig16}, we display a directed acyclic graph, a probabilistic graphical representation of our hierarchical Bayesian model.  Probabilistic graphical models were first used to express hierarchical Bayesian inference with SN Ia by \citet{mandel09,mandel11}.  The probabilistic graphical model describes how the unknown parameters of individual SN Ia (labelled by index $s$) and the hyperparameters of the dust and intrinsic SN Ia populations are related to the measured supernova data $\{\bm{d}_s, z_s\}$, and cosmological parameters $\bm{\Omega}$.  From the intrinsic SN Ia population, described by hyperparameters $\bm{\Theta}_\text{SN}$, a vector of intrinsic light curve parameters $\bm{\psi}_s$ is drawn for each SN $s$.  
Random values of dust reddening $E_s$ and extinction $A_B^s$ for each SN are drawn from the host galaxy dust population, described by hyperparameters $R_B, \tau$.  The intrinsic and dust latent variables combined with distance modulus and light curve fitting error yield the estimated light curve parameters.  The distance modulus is related to the observed redshift through the cosmological parameters and peculiar velocity error. In this paper, we have fixed the cosmological parameters $\bm{\hat{\Omega}}$ for the low-$z$ analysis.

\subsection{Marginal Likelihoods}\label{sec:marglkhd}

A marginal likelihood is obtained from the joint density Eq. \ref{eqn:full_lkhd}.
\begin{equation}\label{eqn:marglkhd}
\begin{split}
P( \bm{d}_s | \, z_s; \bm{\Theta}, \bm{\Omega} ) &= \int \int \int dE_s\, d\mu_s\,  d\bm{\phi}_s\, P( \bm{d}_s, \bm{\phi}_s, E_s, \mu_s | \, z_s; \bm{\Theta}, \bm{\Omega} ) \\
&= | 2 \pi s_E^2 |^{1/2} \times N( \bm{y}_s | \bm{e}_E \hat{E}_s, \bm{\Sigma}_s ) \times  \tau^{-1} \exp\left[\frac{1}{2}\left(\frac{s_E}{\tau}\right)^2 - \frac{\hat{E}_s}{\tau} \right] \times \Phi\left(\frac{\hat{E}_s}{s_E} - \frac{s_E}{\tau}\right)
\end{split}
\end{equation}
where $N( \bm{x}| \, \bm{\mu}, \bm{\Sigma} )$ is a placeholder for $| 2\pi \bm{\Sigma} |^{-1/2} \exp[ -\frac{1}{2} (\bm{x}-\mu)^T \bm{\Sigma}^{-1} (\bm{x}-\mu)]$
 and  we define:
\begin{equation}\label{eqn:invsE}
s_E^{-2} \equiv \bm{e}_E^T \bm{\Sigma}_s^{-1} \bm{e}_E.
\end{equation}
\begin{equation}\label{eqn:Ehat}
\hat{E}_s \equiv s_E^2 \bm{e}_E^T \bm{\Sigma}_s^{-1} \bm{y}_s
\end{equation}
\begin{equation}\label{eqn:y_s}
\bm{y}_s \equiv \bm{d}_s - \bm{\mu}_\psi - \bm{e}_1 \mu_\text{$\Lambda$CDM}(z_s | \, \bm{\Omega})
\end{equation}
\begin{equation}\label{eqn:Sigma_s}
\bm{\Sigma}_s \equiv \bm{\Sigma}_\psi + \bm{W}_s + \bm{e}_1 \bm{e}_1^T \sigma_{\mu|z, s}^2
\end{equation}
where $\bm{\mu}_{\psi}$ and $\bm{\Sigma}_\psi$ are the mean and covariance implied by Eq. \ref{eqn:intrinsic_factorization}.
Assuming conditional independence of the individual SN Ia, the marginal likelihood for the entire sample $\{ \mathcal{D}, \mathcal{Z} \}$ is
\begin{equation}
P( \mathcal{D} | \, \mathcal{Z}; \bm{\Theta}, \bm{\Omega} ) = \prod_{s=1}^{N_\text{SN}} P( \bm{d}_s | \, z_s; \bm{\Theta}, \bm{\Omega} ) .
\end{equation}
This marginal likelihood can be maximized or sampled to estimate the hyperparameters $\bm{\Theta}$ (with $\bm{\Omega}$ fixed):
$P(\bm{\Theta} | \mathcal{D}, \mathcal{Z}, \bm{\Omega}) \propto P( \mathcal{D} | \, \mathcal{Z}; \bm{\Theta}, \bm{\Omega} ) P(\bm{\Theta})$.
Joint cosmological inference would entail computing $P(\bm{\Theta}, \bm{\Omega} | \, \mathcal{D}, \mathcal{Z} ) \propto P( \mathcal{D} | \, \mathcal{Z}; \bm{\Theta}, \bm{\Omega} ) P(\bm{\Theta}) P(\bm{\Omega})$.

\subsection{Photometric Distance Estimates}

If we have point estimates of the hyperparameters $\bm{\hat{\Theta}}$, then the photometric distance modulus conditional on the light curve data (but not the redshift-distance information) is computed from the marginal,
\begin{equation}\label{eqn:photdist}
\begin{split}
P( \bm{d}_s | \,\mu_s; \bm{\hat{\Theta}} ) = \int \, dE_s \,  d\phi_s \, P( \bm{d}_s | \bm{\phi}_s ) \times P(E_s |\, \hat{\tau}) 
\times P( \bm{\psi}_s = \bm{\phi}_s - \bm{e}_1 \mu_s - \bm{e}_E E_s |\, \bm{\hat{\Theta}}_\text{SN}, \hat{R}_B ) .
\end{split}
\end{equation}
The probability density of the photometric distance modulus is $P(\mu_s | \bm{d}_s; \bm{\hat{\Theta}} ) \propto P( \bm{d}_s | \,\mu_s; \bm{\hat{\Theta}} ) P(\mu_s)$, where the prior $P(\mu_s)$ is taken to be flat. Ideally, one would also marginalize over the posterior uncertainty in the hyperparameters $\bm{\Theta}$, but if they are well-determined, this may not be worth the extra effort.

\section{\textsc{Simple-BayeSN} Gibbs Sampling Algorithm}\label{sec:algorithm}

We sketch an MCMC Gibbs sampling algorithm to sample from the global posterior Eq. \ref{eqn:globalposterior} of the \textsc{Simple-BayeSN} hierarchical model.  The purpose of an MCMC algorithm is to generate a sequence of random parameter vectors with a long-run stationary distribution equal to the global posterior. Our Gibbs sampler proceeds by sequentially drawing new parameter values from a full set of conditional posterior densities derived from the global posterior distribution.  Gibbs samplers for hierarchical Bayesian models for SN Ia were previously developed by \citet{mandel09,mandel11,mandel14} and recently by \citet{shariff16}.

We begin a chain with randomized initial values for the apparent colors and distances $\{\bm{\phi}_s, \mu_s\}$, as well as the population hyperparameters $\bm{\Theta}_\text{SN}$, $R_B$, and $\tau$, overdispersed around the maximum likelihood values $\bm{\Theta}_\text{MLE}$.   We alternate between updating the individual SN parameters $\{\bm{\phi}_s, E_s, \mu_s\}$ conditional on the hyperparameters, and updating the population hyperparameters ($\bm{\Theta}_\text{SN}$, $R_B$, $\tau$) conditional on the current values of the set of individual SN parameters.  Steps 1-4 utilize draws from tractable conditional distributions.

\begin{enumerate}

\item Each step below updates the parameters for single SN $s$ given the current values of the hyperparameters.  We cycle through these steps for every SN.

\begin{enumerate}

\item Sample new  dust reddening $\bm{E}_s$ from the conditional posterior, $P( \bm{E}_s |\, \phi_s, \mu_s; \bm{\Theta})$.  

\item Sample new apparent parameters $\bm{\phi}_s$ from the conditional posterior $P(\bm{\phi}_s | \, \bm{E}_s, \mu_s;  \bm{\Theta}, \bm{d}_s, z_s)$.
 
\item Sample a new distance modulus $\mu_s$ from the conditional posterior $P(\mu_s | \, \bm{E}_s, \phi_s;  \bm{\Theta}, z_s)$.
 
\end{enumerate}

\item Sample a new host galaxy reddening scale $\tau$ from $P(\tau| \, \{ E_s\})$.

\item Sample a new value of the dust slope $R_B$ from $P( R_B | \, \{ \phi_s, E_s, \mu_s \}, \bm{\Theta}_\text{SN} )$.

\item Sample the hyperparameters $\bm{\Theta}_\text{SN}$ from $P( \bm{\Theta}_\text{SN} | \, \{ \phi_s, E_s, \mu_s \}, R_B )$.  First, we compute the intrinsic SN Ia parameters $\{ \bm{\psi}_s = \bm{\phi}_s - \bm{e}_1 \mu_s - \bm{e}_E E_s \}$.  Then we sample from the conditional posteriors corresponding to the classical ordinary linear regression problems described by each of Eqns. \ref{eqn:intrinsic1} - \ref{eqn:intrinsic3}.

\item If simultaneously fitting for cosmological parameters, then update $\bm{\Omega}$ from $P(\bm{\Omega} | \, \mu_s, z_s)$ using a Metropolis-Hastings accept/reject algorithm.

\end{enumerate}

We repeat these steps for 5 to 10 $\times 10^3$ cycles, taking several minutes for moderately-sized supernova samples ($N_\text{SN} \sim 300$).  To monitor convergence, we typically run 4-8 parallel chains starting from different initial guesses, and compute the Gelman-Rubin statistic \citep{gelman92}.  The maximum G-R statistic is typically less than 1.02.    We discard the initial 20\% of each chain as burn-in, and concatenate the remaining chains for posterior analysis.

\section{Bayesian Fitting of the Linear Tripp Formula}\label{sec:fit_tripp}

The coefficients of the Tripp formula ($\alpha, \beta$; Eq. \ref{eqn:tripp1}) have been commonly estimated by minimizing a ``$\chi^2$'' of the Hubble residuals.  For example, \textsc{SALT2mu} minimizes a $\chi^2$, modified in the fashion of the FITEXY estimator of \citet{numrec}, to include measurement errors and residual scatter.  However, even with these factors included, the FITEXY $\chi^2$ has known biases, especially when the measurement error or residual scatter in the covariates (e.g. $\hat{c}_s$ or $\hat{x}_s$) is of comparable size to the width of the sample.  This has already been demonstrated in astronomical applications by \citet{bkelly07}, and highlighted for SN Ia analysis with the Tripp formula by \citet{march11}. This frequentist bias is not alleviated even when a log variance normalization factor (dependent on the regression coefficients) is simply added to the $\chi^2$ to make the procedure equivalent to a maximum likelihood.  The bias is minimized when proper hierarchical prior, or population, distributions for the latent covariates are included  \citep{bkelly07}.  Hierarchical models account for the error in the covariates by implementing ``shrinkage'' via population distributions controlled by hyperparameters estimated from the data.  Shrinkage accounts for the fact that the observed distribution of the covariates is wider than the true distribution of the underlying latent variables due to measurement scatter \citep{loredohendry10}.

A hierarchical Bayesian regression model for fitting the conventional Tripp formula can be obtained as a special case of the \textsc{Simple-BayeSN} model.  This case is obtained by ``turning off'' the dust components of the model, $(E_s = \tau = R_B = 0)$, and then re-interpreting the intrinsic (``int'') latent variables and hyperparameters now as apparent (``app'') or extinguished (``ext'') latent variables and hyperparameters.   The intrinsic SN Ia dispersion $\sigma_\text{int}$ is now re-interpreted as the residual scatter $\sigma_\text{res}$ of Eq. \ref{eqn:tripp1}.  This in effect disables the separate modeling of intrinsic and dust components, and instead fits the apparent color-magnitude relation with a single slope $\beta^t_\text{app}$.  With this relabelling (and with $\alpha_c = 0$), this special case is described by  Gaussian population distributions for light curve shape $x_s \sim N(x_0, \sigma_x^2)$, and apparent color $c^\text{app}_s \sim N(c_0, \sigma_{c,\text{app}}^2)$, the latent variable equation (Tripp formula) Eq. \ref{eqn:tripp},  the light curve fitting error likelihood Eq. \ref{eqn:lc_lkhd}, and distance-redshift likelihood Eq. \ref{eqn:zmu_lkhd}.  This special case of the model is equivalent (up to notation) to the hierarchical Bayesian model of \citet{march11}.  Conditional on the cosmological parameters, this is also a special case of the more general structural equation regression model of \citet{bkelly07}.     In this way, we fit the conventional Tripp formula to the data.\\


\bibliographystyle{apj}
\bibliography{apj-jour,sn,stat}{}
\end{document}